\documentclass[prx,superscriptaddress,twocolumn,showkeys]{revtex4-2}

\usepackage{hyperref}
\usepackage{amsfonts, amssymb, amsmath}
\usepackage{graphicx}
	\graphicspath{{figures/}} 
\usepackage[dvipsnames]{xcolor}
\newcommand{\BE}{\begin{equation}}
\newcommand{\EE}{\end{equation}}
\newcommand{\BEA}{\begin{eqnarray}}
\newcommand{\EEA}{\end{eqnarray}}
\newcommand{\on}{\text{on}}
\newcommand{\off}{\text{off}}

\newcommand{\z}{z}
\newcommand{\zbar}{\bar z}
\newcommand{\Z}{{\cal Z}}
\newcommand{\Zbar}{\bar {\cal Z}}

\newcommand{\act}{\text{act}}

\newcommand{\step}{\text{step}}

\renewcommand{\a}{{\bf a}}
\renewcommand{\b}{{\bf b}}
\renewcommand{\d}{{\rm d}}

\newcommand{\ep}{\epsilon}
\newcommand{\lam}{\lambda}

\newcommand{\x}{{\bf r}}

\renewcommand{\epsilon}{\varepsilon}
\begin{document}


\title{Nonequilibrium antigen recognition during infections and vaccinations}


\author{Roberto Morán-Tovar}
\author{Michael Lässig}
\email[Corresponding author: ]{mlaessig@uni-koeln.de}
\thanks{}
\affiliation{Institute for Biological Physics, University of Cologne, Z\"ulpicherstr.~77, 50937 K\"oln, Germany}



\begin{abstract}
The immune response to an acute primary infection is a coupled process of antigen proliferation, molecular recognition by naive B cells, and their subsequent proliferation and antibody shedding. This process contains a fundamental problem: the recognition of an exponentially time-dependent antigen signal. Here we show that B cells can efficiently recognise new antigens by a tuned kinetic proofreading mechanism, where the molecular recognition machinery is adapted to the complexity of the immune repertoire. This process produces potent, specific and fast recognition of antigens, maintaining a spectrum of genetically distinct B cell lineages as input for affinity maturation. We show that the proliferation-recognition dynamics of a primary infection is a generalised Luria-Delbrück process, akin to the dynamics of the classic fluctuation experiment. This map establishes a link between signal recognition dynamics and evolution. We derive the resulting statistics of the activated immune repertoire: antigen binding affinity, expected size, and frequency of active B cell clones are related by power laws, which define the class of generalised Luria-Delbr\"uck processes. Their exponents depend on the antigen and B cell proliferation rate, the number of proofreading steps, and the lineage density of the naive repertoire. We extend the model to include spatio-temporal processes, including the diffusion-recognition dynamics of a vaccination. Empirical data of activated mouse immune repertoires are found to be consistent with activation involving about three proofreading steps. The model predicts key clinical characteristics of acute infections and vaccinations, including the emergence of elite neutralisers and the effects of immune ageing. More broadly, our results establish infections and vaccinations as a new probe into the global architecture and functional principles of immune repertoires.

\end{abstract}


\maketitle

\section{Introduction}

\noindent B cells are a central part of the human adaptive immune system. These cells recognise pathogens by specific binding: B cell receptors (BCRs) located in the cellular membrane bind to antigenic epitopes, which are cognate binding loci on the surface of pathogens. To capture a wide range of a priori unknown pathogens, humans produce a large and diverse naive B cell repertoire, estimated to contain about $L_0 \sim 10^{9}$ lineages with distinct BCR genotypes~\cite{Altan-Bonnet2020, Glanville2009, Elhanati2015} and a larger but comparable number of circulating naive B cells~\cite{Altan-Bonnet2020, Morbach2010}. Acute infections and vaccinations with a live-attenuated virus are characterised by rapid, initially exponential growth of the pathogen population. An infection often starts with few particles and reaches peak densities of order $10^{8} {\rm ml}^{-1}$ within a few days~\cite{Smith2010, Pawelek2012, Goyal2021, Sender2021}. 
Inactivated antigens, as used in most influenza vaccines, generate a similar signal increase by diffusion from the point of vaccination to a point of recognition, which is typically in a lymph node. At some stage of this process, free antigens start to bind to B cells in lineages of sufficiently high binding affinity. Antigen binding can activate B cells, triggering rapid proliferation and shedding of free antibodies (membrane-detached BCRs) that eventually clear the pathogen. Activated B cells also form germinal centres and create immunological memory~\cite{Goodnow2010, Lam2019}. Activated repertoires are estimated to contain multiple B cell lineages, $L_\act \gtrsim 10^2$ in mouse models~\cite{Tas2016, Mesin2020}.

The exponential growth of the antigenic signal, together with a large number of circulating B cell lineages, presents a formidable real-time specificity problem for recognition. Consider an antigen that activates a high-affinity B cell lineage at a given point of time. One day later, at a $>100$fold higher density, the antigen can potentially activate a large number of low-affinity lineages, generating a poor overall response of the immune repertoire. The actual process activates only a tiny fraction of the pre-infection repertoire. Previous work has established upper bounds of order $L_\act / L_0 \lesssim 10^{-5}$~\cite{Cancro1978, Perelson1979}, and a lower bound follows from recent data, $L_\act / L_0 \gtrsim 10^{-7}$~\cite{Tas2016, Mesin2020}. How is such highly specific immune response possible? At its core, this is a fundamental problem of signal recognition: how to process an exponentially increasing signal, here of a growing antigen population, for a fast and specific response. Molecular recognition problems, in and outside the context of immunology, have so far been studied mostly for steady-state signals~\cite{Hopfield1974, Ninio1975, Mckeithan1995, Gromadski2004, Johansson2012}. In this paper, we show that optimal recognition of an exponential signal depends on three {\em a priori} unrelated factors: the signal processing mechanism, the growth rates of signal and response, and the complexity of the recognition repertoire. 

\begin{figure*}[ht!]
\centering
\includegraphics[width= 0.9 \textwidth]{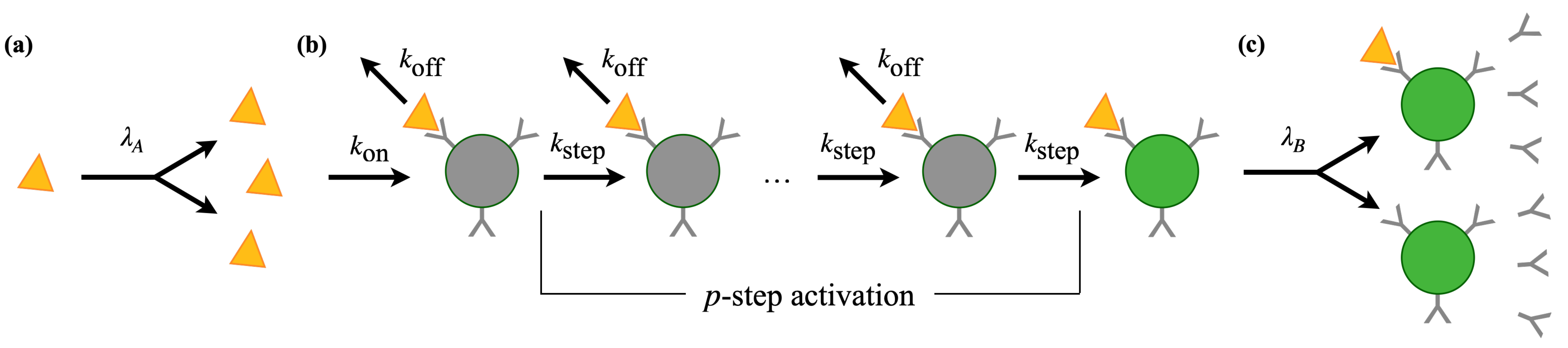}
\caption{ {\bf Antigen recognition in a primary infection.} 
In a minimal model, immune recognition of a new antigen involves three stages.
{\bf (a)}~Exponential replication of free antigens with rate $\lam_A$.
{\bf (b)}~Activation of naive B cells: binding of antigens with association rate $k_\on$, dissociation rate $k_\off$, and $p$ irreversible activation steps with rate $k_\step$. This mechanism yields a total activation rate $u_\act$ that decays as $u_\act \sim (k_\step/k_{\rm off})^{p}$ in the low-affinity regime (see text). 
{\bf (c)}~Exponential replication of activated B cells and proliferation of free antibodies with rate $\lam_B$.}
\label{fig:1}
\end{figure*}

Our basic model for the recognition dynamics has three steps: (i) antigen proliferation, (ii) nonequilibrium molecular recognition and cell activation, and (iii) subsequent proliferation of activated B cells [see Fig.~\ref{fig:1}]. An extension of the model includes antigen diffusion, which turns out to be a relevant factor for the recognition dynamics of vaccinations. The B cell activation process contains kinetic proofreading: a series of multiple, thermodynamically irreversible steps~\cite{Hopfield1974, Ninio1975}. In an appropriate regime of rate parameters, processes with kinetic proofreading are known to increase the affinity discrimination of their output compared to near-equilibrium processes. For the activation process of Fig.~\ref{fig:1}, we show that the activation rate of weak binding B cells depends on the antigen-BCR binding constant and on the number of activation steps, $u_\act \sim K^{-p}$. 

Kinetic proofreading has been recognised as a key step in the activation of T cell immunity~\cite{Mckeithan1995, Goldstein2004, Francois2013, Pettmann2021}. For B cells, evidence for proofreading comes from experimental observations of characteristic time lags in activation, but little is known about the underlying molecular mechanism~\cite{Tsourkas2012}. Observed mechanisms of immune cell activation by membrane-bound antigens include BCR clustering, membrane spreading-contraction, quorum sensing, and molecular tug-of-war extraction forces~\cite{Harwood2008, Yang2010, Butler2013, Jiang2023}. Such mechanisms may contribute to proofreading, but their relevance for the specificity of {\em initial} antigen recognition addressed in this article remains unclear. Here we use a minimal $p$-step model of activation to show that proofreading is essential for specific and timely recognition of an exponential antigen signal. This result complements previous work on steady-state signal recognition by kinetic proofreading~\cite{Hopfield1974, Ninio1975, Mckeithan1995, Gromadski2004, Johansson2012}.

To understand how activation and proofreading act in the face of an exponentially increasing input signal, we treat the recognition dynamics as a generalised Luria-Delbr\"uck process. This process resembles the proliferation-mutation dynamics of the classical Luria-Delbr\"uck experiment~\cite{Luria1943}: the antigen corresponds to the wild-type, activation to mutation, and B cell lineages to mutant cell lineages. The new feature of the infection or vaccination dynamics, which has no analogue in the original Luria-Delbr\"uck process, is that each B cell lineage has a specific antigen binding constant $K$. This sets the density of B cell lineages available for activation, $\Omega_0 \sim K^{\beta_\act}$, and modulates their activation rate. 

In the first part of the paper, we develop the theory of generalised Luria-Delbr\"uck processes for antigen recognition, and we compute the distribution of lineages in activated B cell repertoires. In the second part, we turn to biological implications of the recognition dynamics. 
Our model predicts optimal immune responses to an exponentially increasing pathogen population, tuned to a balance between speed and potency, at an intermediate number of proofreading steps. Recent data of vaccination-activated mouse immune repertoires~\cite{Tas2016, Mesin2020} are shown to be consistent with this prediction. The model further predicts that activated immune repertoires of different hosts responding to the same antigen show giant fluctuations, similar to mutant populations in a classical fluctuation experiment. Such fluctuations are a hallmark of Luria-Delbr\"uck processes~\cite{Luria1943, Lea1949, Mandelbrot1974, Kessler2013, Zheng1999}. In a primary immune response, giant fluctuations are generated by ``jackpot'' clones of large size and high antigen affinity. We derive the underlying statistics of activated repertoires and infer clinically important characteristics of primary infections and vaccinations.


\section{Theory of exponential antigen recognition}

In this section, we first derive the activation probability of individual B cell lineages in an acute infection, which depends on their antigen binding kinetics and on the growth rate of the pathogen. In a second step, we compute the activation characteristics of the entire B cell repertoire, which also depends on the density of lineages and determines the physiological immune response to an infection. The underlying basic model of antigen proliferation, molecular recognition, and subsequent proliferation of activated B cells describes antigens and B cells by homogeneous densities, neglecting their spatial structure. At the end of the section, we develop a fully spatio-temporal model of the recognition dynamics, which includes antigen diffusion from the starting point of the infection to a locus of recognition. This model shows that generalized Luria-Delbr\"uck processes apply to infections and vaccinations, validating the homogeneous-system approximation for infection responses and establishing a new, diffusion-limited regime relevant for vaccination responses.

\begin{figure}[h!]
\begin{center} 
\includegraphics[width=.86\columnwidth]{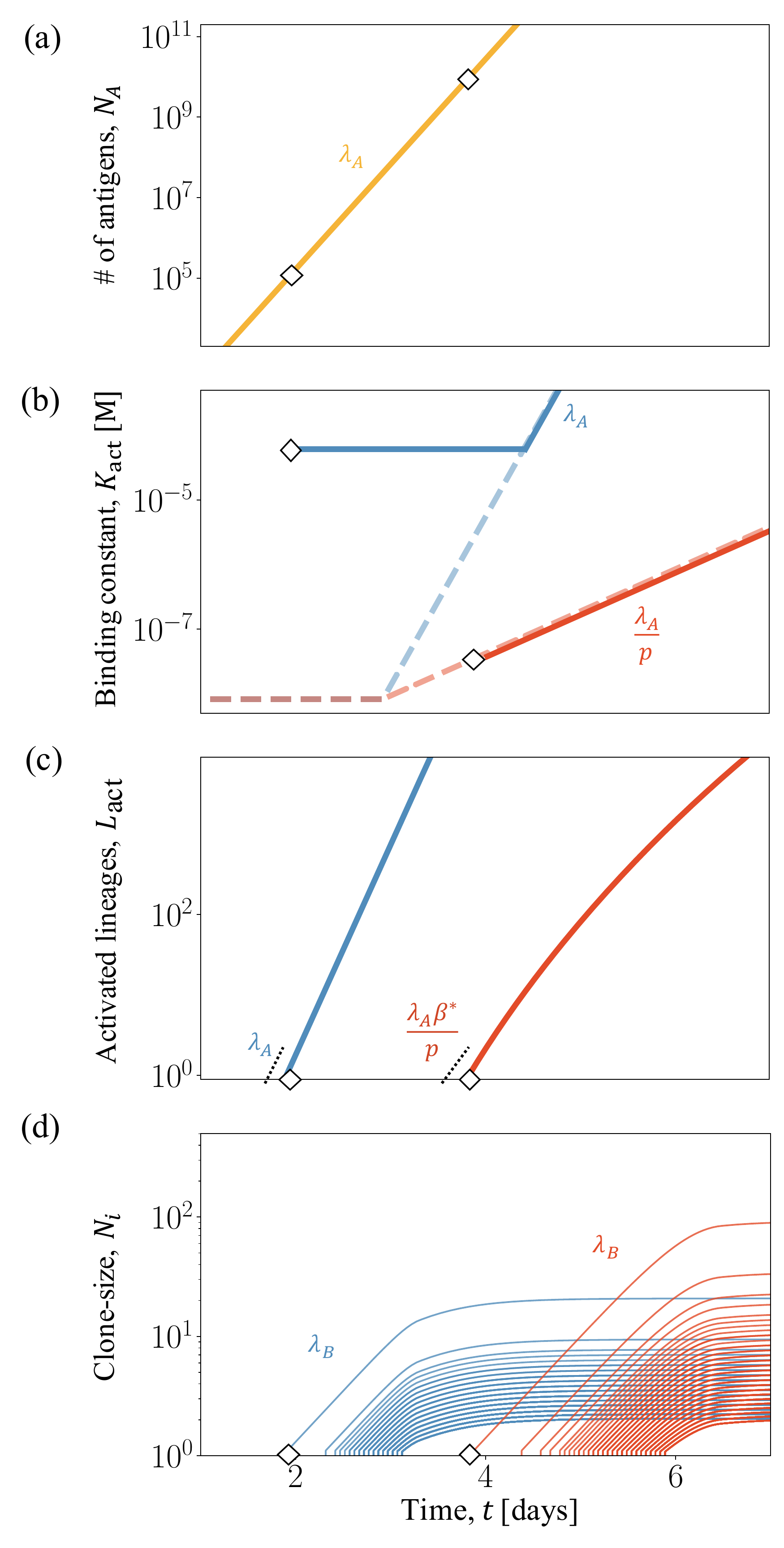}
\end{center}
\vspace*{-0.5cm} 
\caption{ {\bf Generalised Luria-Delbr\"uck replication-activation dynamics.} 
{\bf (a)} The antigen population, $N_A(t)$, grows exponentially with rate $\lam_A$. Diamonds mark the start of B cell activation in the low-specificity (LS) and the high-specificity (HS) regime. 
{\bf (b)} The average binding constant of activation, $K_\act(t)$, is initially constant (LS, blue) or increases with rate $\lam_A / p$ (HS, red). 
{\bf (c)} The number of activated lineages, $L_\act (t)$, grows exponentially with initial rates $\lam_A$ (LS) and $\beta_\act \lam_A/p$ (HS). 
{\bf (d)} The population of activated lineages, $N_i (t)$ ($i = 1,2, \dots, $), shows initially exponential growth with rate $\lam_B$. In the mapping to a classical Luria-Delbr\"uck process, the antigen population corresponds to the wild-type cell population, activation to mutation, and B cell linages to mutant cell clones under selection. The generalised Luria-Delbrück process couples growth to neutralisation function and repertoire complexity, setting independent growth rates of $K(t)$ and $L(t)$. 
Analytical results are shown for the following parameters: activation steps, $p =1$ (LS), $p = 4$ (HS); kinetic parameters, $k_\on = 10^6M^{-1}s^{-1}$, $k_\step =0.5\text{min}^{-1}$; number of BCRs per cell, $b_0=10^5$; growth rates $\lam_A=6 \text{d}^{-1}$, $\lam_B=2 \text{d}^{-1}$; repertoire size $L_0= 10^9$; carrying capacity $\bar N =10^4$.
 }
\label{fig:2}
\end{figure}

\vspace{1ex} \paragraph*{\bf Antigen-BCR binding.} 
In the initial phase of an infection, the antigen population grows exponentially with a rate $\lambda_A$ [see Figs.~\ref{fig:1}(a) and~\ref{fig:2}(a)]. For viral pathogens, this  process starts with few localized antigen copies and reaches population numbers $N_A(t)$ of order $10^{12}$ within about 5 days, which implies replication factors $>100$/d~\cite{Smith2010, Pawelek2012, Goyal2021, Sender2021}. In a homogeneous system, 
the total association rate between free antigens and circulating naive B cells of a given lineage is given by 
\BE
u_{\on}(t) = N_A(t) b_0 k_\on \rho_B,
\label{u_on} 
\EE
where $N_A(t)=\exp{(\lam_At)}$ is the total number of antigen particles, $b_0 \sim 10^5$ is the number of BCR per B cell~\cite{Altan-Bonnet2020}, $\rho_B \sim 1 \mathrm{cell/L}$ is the density of B cells per lineage~\cite{Altan-Bonnet2020, Glanville2009, Elhanati2015, Morbach2010}, and $k_\on$ is the molecular association rate [see Fig.~\ref{fig:1}(b)]. Association is known to be diffusion-limited with typical rates $k_\on \sim 10^6~\mbox{M}^{-1} \mbox{s}^{-1}$~\cite{Pecht1972}. Therefore, differences in antigen binding affinity between different B cell lineages result primarily from differences in the dissociation rate $k_\off$. Human B cells have dissociation rates in the range $k_\off = 10^{-5}-10^{1} \mbox{s}^{-1}$~\cite{Altan-Bonnet2020}; the corresponding equilibrium binding constant, $K = k_\off / k_\on$, varies in the range $K = 10^{-11}-10^{-5} \mbox{M}$. This constant is related to the lineage-specific energy gap between the bound and the unbound state, $K = K_0 \exp(\Delta E)$, where $K_0$ is a normalisation constant and all energies are measured in units of $k_BT$ [see Appendix~\ref{app:A}]. Importantly, with these parameters, the fraction of antigen-bound B cells remains small throughout the infection process. Antigen consumption by B cells, which becomes relevant during affinity maturation, can be neglected at this stage of the process. 

\vspace{1ex} \paragraph*{\bf B cell lineage activation.}

Upon binding, we assume that B cells undergo a series of $p$ thermodynamically irreversible steps to activation [see Fig.~\ref{fig:1}(b)]. That is, cells in each intermediate state transform to the next state (with rate $k_\step$) or unbind from the antigen (with rate $k_\off$), but do not revert to the previous state. The stepwise, stochastic activation dynamics is an inhomogeneous Poisson process, the output of which is the activation rate. In the relevant regime of low antigen concentration, the activation rate per B cell lineage, $u_\act (t)$, takes the form
\BE
u_\act(K, t) = \left(1+\frac{K}{K_\step}\right)^{-p} u_\on(t)
\label{eq:u_act}
\EE
with $K_\step = k_\step/k_\on$. In the low-affinity regime, this rate has the asymptotic form $u_\act (K, t) \sim (K/K_\step)^{-p} \sim (k_\step/k_{\rm off})^{p}$, which can be read off from Fig.~\ref{fig:1}(b): each activation step generates a factor $(k_\step/k_{\rm off})$ relating the thermodynamic weights of consecutive intermediate states. 

Next, we compute the activation probability of a B cell lineage up to time $t$, $R(K, t)$. For exponential antigen growth, we find 
\BE
R(t, K) = 1-  \exp{\left[-\frac{u_\act(t, K)}{\lam_A}\right]}; 
\label{eq:P_act}
\EE
[see Appendix~\ref{app:A}]. This equation describes a moving front of deterministic activation, $K_1 (t) = K_\step \exp[ (\lam_A/p) (t - t_1)]$, where $R$ reaches values of order 1 [see Fig.~S1]. The front starts at affinity $K_\step^{-1}$ and time 
$t_1 = \log [\lambda_A/(b_0 k_\on \rho_B)]/\lambda_A$. With increasing antigen concentration, it moves towards lineages of decreasing antigen affinity at a $p$-dependent speed [see Fig.~\ref{fig:2}(b)]. Ahead of the front, for $K \gg K_1 (t)$, activation of individual lineages is a rare stochastic event. For $p =1$, activation is asymptotically proportional to the inverse equilibrium constant, or Boltzmann factor, $R \sim K^{-1} \sim \exp(- \Delta E)$. For $p > 1$, the non-equilibrium dynamics of kinetic proofreading generate stronger suppression of activation for weak binders, $R \sim K^{-p}$~\cite{Hopfield1974, Ninio1975}. Kinetic proofreading appears to be the simplest mechanism to generate deterministic activation of high-affinity lineages together with strong suppression of low-affinity lineages; mechanisms with reversible antigen-receptor binding have $R \sim K^{-1}$ or remain in the stochastic regime ($R \ll 1$) under the physiological conditions of an early primary infection [see Fig.~S1 and Appendix~\ref{app:A}]. 

Lineage activation marks the onset of the immune response to a new antigen. Activated cells proliferate exponentially with an initial rate $\lambda_B$ that is comparable to $\lambda_A$~\cite{Bocharov1994} and shed free antibodies that can neutralise antigens [see Fig.~\ref{fig:1}(c)]. As the activated repertoire grows, cells start to compete for space and resources, including T cell help~\cite{Schwickert2011}. Here we model the clone dynamics as logistic growth, 
\BE
\dot N_j (t) = \lambda_B N_j (t) \bigg ( 1 - \sum_j \frac{N_j (t)}{\bar N} \bigg ),
\label{eq:N_b}
\EE
where $N_j (t)$ is the number of activated cells in clone $j$ and $\bar N \sim 10^4$~\cite{Bocharov1994} is a carrying capacity for the total size of the activated repertoire [see Fig.~\ref{fig:2}(d)]; here and below, overbars refer to the repertoire statistics at carrying capacity.

\begin{figure}[t]
\centering
\includegraphics[width=1\columnwidth]{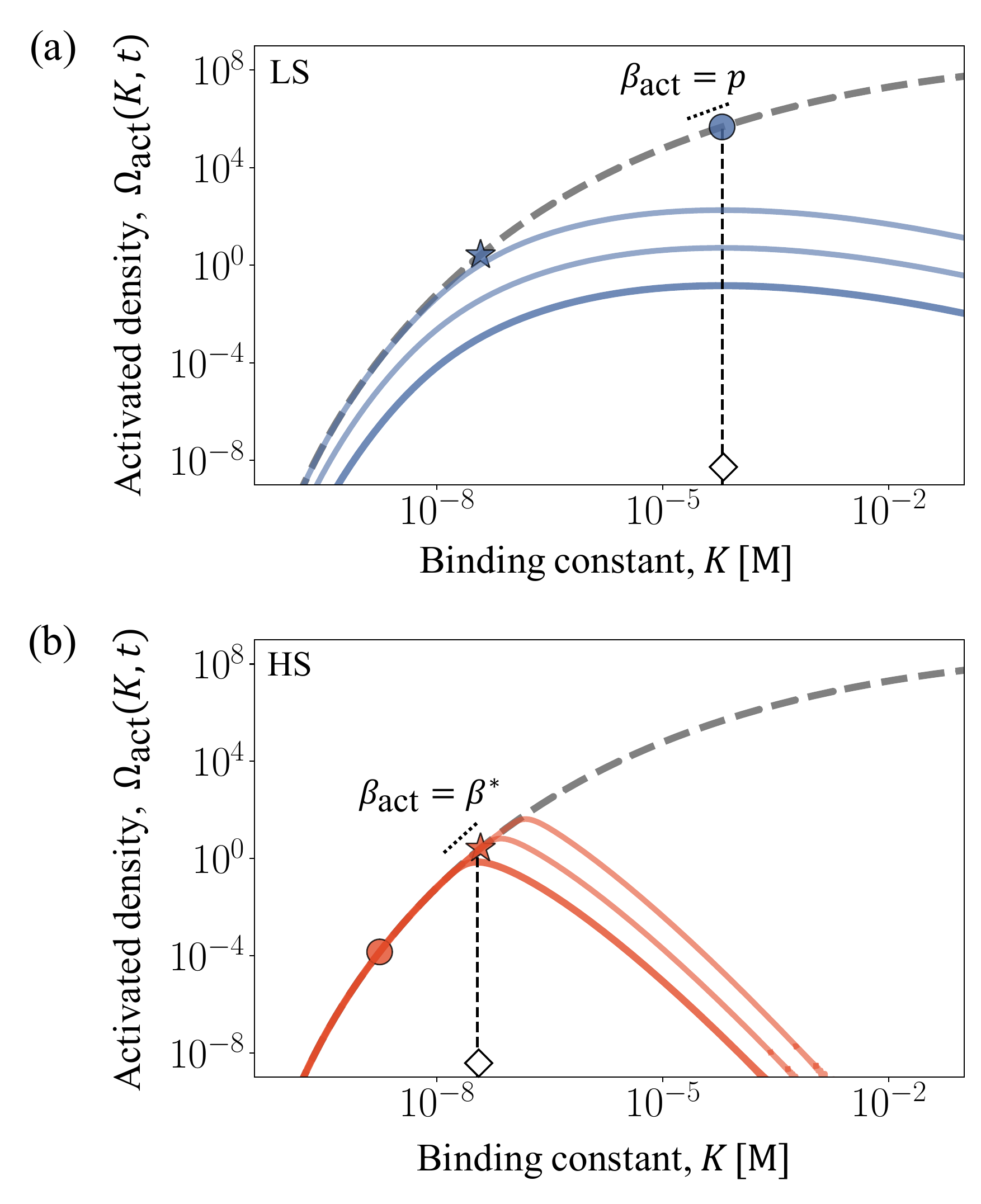}
\caption{ {\bf B cell response repertoires.} 
Density of naive B cell lineages, $\Omega_0 (K)$ (gray); density of activated lineages, $\Omega (K, t)$, at the start of activation($t = t_\act$, thick) and at two subsequent time points (thin) in the {\bf (a)} LS regime ($p =1$) and {\bf (b)} HS regime ($p = 4$). Specific values of $K$ are marked on the spectral function: $K_p$ (dot); strongest antigen binding in typical individuals, $K^*$ (star); start of activation, $K_\act$ ({open diamond}) with initial inverse activation temperature, $\beta_\act$ (dotted line). Energy model: TCRen; other parameters as in Fig.~\ref{fig:2}.
\label{fig:3}
}
\end{figure}

\vspace{1ex} \paragraph*{\bf Repertoire response to a given antigen.}
To characterise the immune repertoire available for primary response against a given antigen, we grade naive B cell lineages by their affinity to the antigen's binding locus (epitope). We use a simple sequence-specific binding energy model, where epitopes and their cognate BCR are sequence segments, $\a = (a_1, \dots, a_\ell)$ and $\b = (b_1, \dots, b_\ell)$. Binding aligns these segments and couples pairs of aligned amino acids, and the binding energy gap $\Delta E (K) = \log (K/K_0)$ is additive, 
\BE
\Delta E (\a, \b) = \sum_{k = 1}^\ell \varepsilon (a_k, b_k). 
\label{E} 
\EE
For a given antigen, this model defines the density of naive lineages available for activation, $\Omega_0 (K) = (K \d / \d K) L_0 (K)$, where $L_0 (K)$ is the expected number of lineages in an individual with binding constant $<K$ to the epitope $\a$ [see Fig.~\ref{fig:3} and Appendix~\ref{app:B}]. Here we assume that naive repertoires are randomly sampled from an underlying amino acid distribution~\cite{Elhanati2015}. Hence, most lineages bind only weakly to a new antigen ($K \sim 0$). The expected minimum binding constant in an individual, $K^*$, is given by the extreme value condition $L_0 (K^*) = 1$. This point is to be distinguished from the global minimum for a given antigen, $K_m$, which often corresponds to a unique BCR genotype $\b_m$ (called the Master sequence). Because individual repertoires cover only a small fraction of the BCR genotype space, the expected maximum antigen affinity remains substantially below the Master sequence ($K^* > K_m$). To characterise the strong-binding tail of the lineage spectrum, we define the micro-canonical entropy, $S(K) = \log \Omega_0 (K) + {\rm const.}$, and the associated inverse reduced temperature, 
\BE
\beta (K) = \frac{{\rm d} S(K)}{{\rm d} (\Delta E (K))}.
\label{eq:beta}
\EE
This function measures the exponential increase in lineage density with binding energy, $\Omega_0 (K') \sim \exp[\beta (K) \, \Delta E (K')]$, in the vicinity of a given point $K$. As we will show below, the inverse temperature at maximum binding, $\beta^* = \beta (K^*)$, is a key determinant of the activation dynamics. 

\vspace{1ex} \paragraph*{\bf Repertoire response to diverse antigens.} 
How comparable are the response repertoires of different antigens? To address this question, we evaluate the lineage spectrum $\Omega_0 (K)$ for a random sample of antigenic epitopes $\a$ [see Fig.~S2]. The interaction energy matrix $\ep (a, b)$ of our main analysis is proportional to the TCRec matrix $\ep (a, b)$ originally inferred for T cell receptors \cite{Karnaukhov2022}; similar spectra are obtained from the Miyazawa-Jernigan matrix~\cite{Miyazawa1996} and from normally distributed random energies. For a given antigen, the spectral density depends on broad statistical features of the energy matrix, including the binding length $\ell$ and the variance of interaction energies, $\sigma^2_\epsilon$ [see Appendix~\ref{app:B}]. Here we determine these parameters from observed binding constants $K^* \sim 10^{-7}$M and $K_m \sim 10^{-11}$M of high-affinity antibodies generated in primary infections and of ultra-potent antibodies, respectively~\cite{Eisen2014, Altan-Bonnet2020}. 

Remarkably, these physiological constraints generate a consistent ensemble of response repertoires [see Fig.~S2]. First, the distributions of inferred binding lengths and of the rms. energy variation per site are strongly peaked around values $\ell \sim 20$ and $\sigma_\epsilon / (\ell^{1/2}) \sim 1$, respectively, which are in tune with known examples. Second, the lineage densities $\Omega_0 (K)$ depend only weakly on the antigen sequence $\a$ and have a nearly universal shape determined solely by the overall repertoire size $L_0$. In other words, the antigen-averaged spectral density $\Omega_0 (K)$ captures the response repertoire available in a typical primary infection. In particular, for a given value of $L_0$, response repertoires of different antigens with similar $K^*$ and $K_m$ have similar inverse activation temperatures, $\beta^* = (2.5 \pm 0.3)$ for humans ($L_0 = 10^9$) and  $\beta^* = (2.2 \pm 0.3)$ for mice ($L_0 = 10^8$)~\cite{Altan-Bonnet2020, Glanville2009, Elhanati2015}.

\vspace{1ex} \paragraph*{\bf Repertoire activation.} 
The spectral density of naive lineages and the recognition function $R(K, t)$ determine the time-dependent density of activated lineages, 
\BE
\Omega_\act (K, t) = \Omega_0 (K) \, R(K,t). 
\label{eq:rho_act} 
\EE
In Fig.~\ref{fig:3}, we plot $\Omega_\act (K, t)$ at subsequent times for two different numbers of activation steps, with and without proofreading ($p = 1, 4$).  
The spectral density of activated lineages is strongly peaked; its high-affinity flank is given by the density of naive lineages, its low-affinity flank by the proofreading-dependent activation dynamics. This function determines two repertoire summary statistics: the expected number of activated lineages, $L_\act (t) = \int \Omega_\act (K,t) \, \d K/K$, and their average 
binding constant $K_\act (t) = \int K \Omega_\act (K,t) \, \d K/K$. Activation starts at an expected time $t_\act$ given by the condition $L_\act (t_\act) = 1$. This sets the initial binding constant $K_\act \equiv K_\act (t_\act)$ [marked by diamonds in Fig.~\ref{fig:3}] and the inverse activation temperature $\beta_\act \equiv \beta_\act (K_\act)$ [marked by a tangent dotted line in Fig.~\ref{fig:3}]. Importantly, activation has two dynamical regimes. 

In the {\em low-specificity} (LS) regime, for small values of $p$, activation is peaked on the low-affinity flank of the spectral function [see Fig.~\ref{fig:3}(a)]. The onset time $t_\act$ and the starting point $K_\act = K_p$ are determined by the condition $\beta_\act = p$, which follows from the asymptotic form $R \sim K^{-p}$ given by Eq.~(\ref{eq:P_act}). In this regime, the number of activation steps, $p$, determines the specificity of recognition; the activation probability and clone size of individual lineages remains small. The LS activation dynamics is characterised by 
\BE
\begin{array}{rcl}
\beta_\act & = & p, 
\\
t_\act & = & t_1 + \frac{1}{\lam_A}\log{\left[ \left (\frac{K_p}{K_\step} \right )^p \Omega_0^{-1} (K_p) \right]}, 
\\
 K_\act (t) & = & K_p, 
 \\
 L_\act (t) & = & \exp\left[ \lam_A (t - t_\act)\right],   
 \end{array}
 \qquad \mbox{(LS)}
 \label{eq:act1}
 \EE
as shown in Fig.~\ref{fig:2}(b) and~\ref{fig:2}(c) [see Appendix~\ref{app:B}]. This regime ends at a crossover point $p = \beta^*$, where $K_p$ reaches the expected minimum binding constant, $K^*$. 

In the {\em high-specificity} (HS) regime, for $p > \beta^*$, activation starts at a later time, $t_\act = t^*$, and at repertoire parameters $K_\act = K^*$ and $\beta_\act = \beta^*$, then follows the deterministic front $K_\act (t) = K_1 (t)$ (Fig.~\ref{fig:3}B). Hence, lineages are activated deterministically and in order of decreasing antigen affinity. In this regime, the spectral density of the naive B cell repertoire determines the specificity of recognition; high-affinity lineages reach substantial clone size. 
We find in the HS activation dynamics 
 \BE
\begin{array}{rcl}
\beta_\act & = & \beta^*,
\\
t_\act & = & t^* = t_1 + \frac{1}{\lam_A} \log{\left[ \left (\frac{K^*}{K_\step} \right )^p \right]}
\\
 K_\act (t) & = & K^* \exp\left[ \frac{\lam_A}{p} (t - t^*)\right], 
 \\
 L_\act (t) & = & \exp\left[ \frac{\lam_A \beta^*}{p}(t - t^*)\right]
 \end{array}
 \qquad \mbox{(HS)}
 \label{eq:act2}
 \EE
as shown in Fig.~\ref{fig:2}(b) and~\ref{fig:2}(c) [see Appendix~\ref{app:B}]. In the next section, we will show that these regimes generate drastically different immune responses.

\begin{figure*}[ht!]
\centering
\includegraphics[width= 1\textwidth]{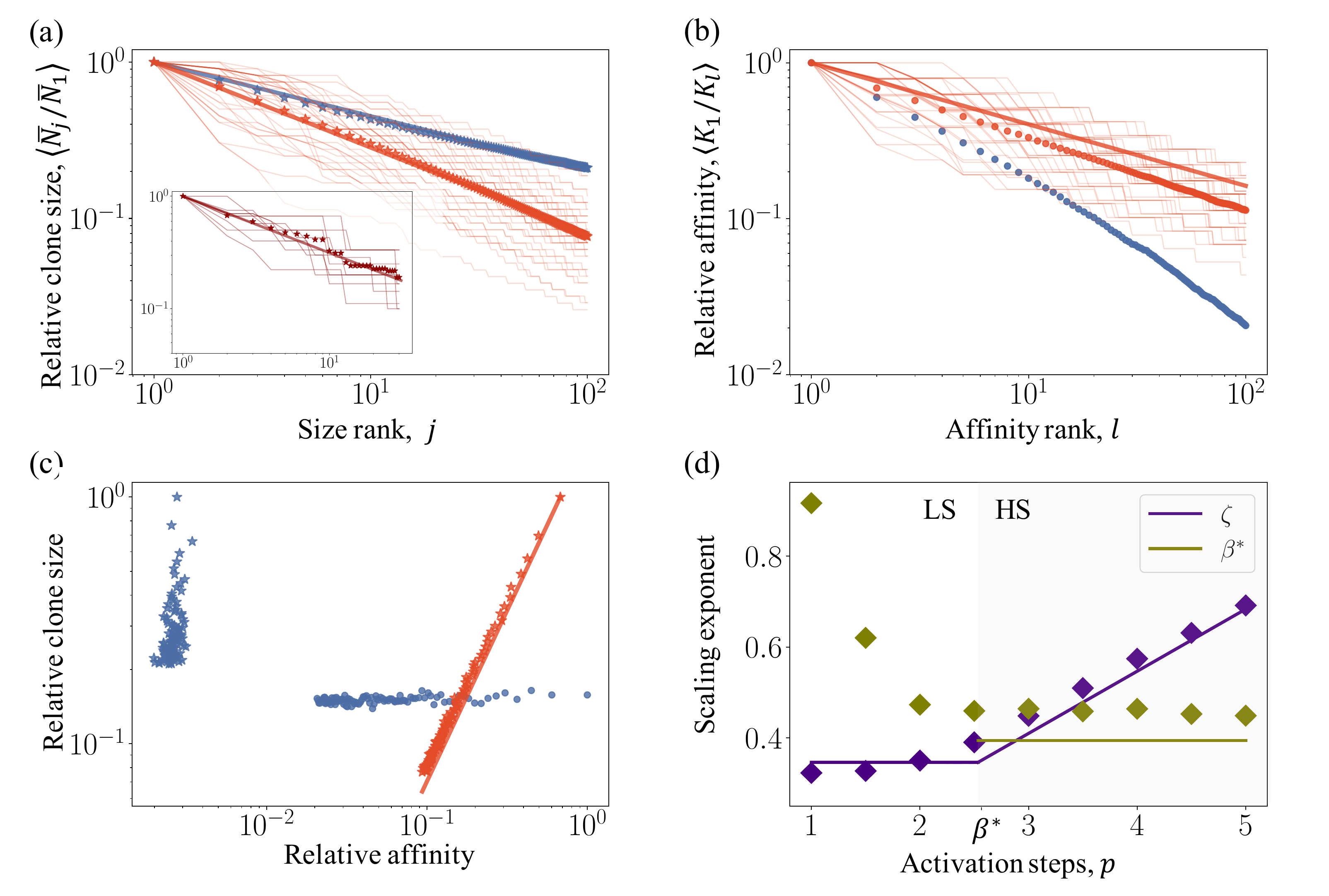}
\caption{ {\bf Clone size and affinity statistics of activated repertoires. }  
{\bf (a)} Clone size statistics. Simulation results for the average relative clone size, $\langle \bar N_j/\bar N_1 \rangle$, are plotted against the size rank, $j = 1, \dots 100$, for $p = 1, 4$ (blue and red, respectively). Lines mark the expected power law, $\langle \bar N_j \rangle \sim j^{-\zeta}$, as given by Eq.~(\ref{eq:x_i}). Thin lines give rankings in a set of randomly chosen individuals. 
Insert: empirical rank-size relation inferred from the repertoire data of ref.~\cite{Tas2016, Mesin2020}.
{\bf (b)} Affinity statistics. Simulation results for the average relative antigen binding constant, $\langle K_l/K_1\rangle$, are plotted against the affinity rank, $l = 1,\dots, 100$. A line marks the expected power law in the HS regime, $\langle K_l \rangle \sim l^{1/\beta_\act}$, as given by Eq.~(\ref{eq:e_i}). 
{\bf (c)} Size-affinity correlation. HS regime: average relative affinity, $\langle K_j/K_1\rangle$ vs.~average relative clone size, $\langle \bar N_j/\bar N_1\rangle$ for the largest clones ($j = 1, \dots, 100$; red stars), together with the expected power law correlation, $\langle \bar N_j \rangle \sim \langle K_j\rangle ^{-\zeta\beta^*}$ (red line). LS regime: $\langle K_j/K_1\rangle$ vs.~$\langle \bar N_j/\bar N_1\rangle$ for the largest clones ($j = 1, \dots, 100$; blue stars), and $\langle K_l/K_1\rangle$ vs.~$\langle \bar N_l/\bar N_1\rangle$ for high-affinity clones ($l = 1, \dots, 100$; blue circles), indicating loss of the size-affinity correlation. 
{\bf (d)} Empirical repertoire exponents from size and affinity rankings are plotted as functions of the number of activation steps, $p$ (diamonds); the high-specificity regime is marked by shading. Lines mark power laws emerging from the Luria-Delbr\"uck proliferation-activation dynamics; their exponents $\zeta$ and $1/\beta^*$ depend on the proliferation rates of antigen and activated B cells, the activation temperature, and the number of proofreading steps. 
Model parameters as in Fig.~\ref{fig:2}.
}
\label{fig:4}
\end{figure*}

\vspace{1ex} \paragraph*{\bf Clone size and affinity statistics.}
B cell immune repertoires are known to have broad variation of clone sizes, which can be described by power-law distributions~\cite{Weinstein2009, Desponds2016, Chardes2022}. The proliferation-activation process of acute infections provides a simple explanation for such power laws: it relates observables that depend exponentially on time [see Fig.~\ref{fig:2}]. First, consider the relation between clone size and probability of occurrence in an individual's repertoire. More lineages are activated later ($L \sim \exp[(\lambda_A \beta_\act / p) \, t]$), but these clones reach smaller size ($\bar N \sim \exp[- \lambda_B t]$). This relates clone size to rank, 
\BE
\langle \bar N_j \rangle \sim j^{-\zeta}
\label{eq:x_i}
\EE
with 
\BE
\zeta= \frac{\lam_B p}{\lam_A\beta_\act}
\label{eq:zeta}
\EE
and $\beta_\act = \min (p, \beta^*)$ [see Appendix~\ref{app:B}]. In what follows, we refer to $\zeta$ and $\beta_\act$ as the {\em size exponent} and the {\em phenotype exponent} of the recognition dynamics, respectively. The index $j = 1,2, \dots$ again orders the clones in an individual's repertoire by size. The cumulative distribution aggregated over individuals has the form {$\Phi (N) \sim N^{-1/\zeta}$}, which is equivalent to Eq.~(\ref{eq:x_i}), and spans 3 orders of magnitude in size [see Fig.~S3 and Appendix~\ref{app:B}]. Simulations confirm these power laws; the clone-rank statistics in a set of randomly chosen individuals follows the same pattern [see Fig.~\ref{fig:4}(a) and~S2]. In the HS regime, the exponent $\zeta$ increases monotonically as a function of $p$, reflecting the increasing bias to large clone size generated by proofreading [see Fig.~\ref{fig:4}(d)].
The activation dynamics of a primary infection include a recognition phenotype (here, antigen affinity), generating additional power laws observable in repertoire data. In the HS regime, activation occurs on a moving front, as given by Eq.~(\ref{eq:act2}). This relates affinity to rank, 
\BE
\langle K_{l} \rangle \sim l^{1/ \beta^*}
\qquad \mbox{(HS)}, 
\label{eq:e_i}
\EE
where the index $l$ orders clones by decreasing affinity [see Fig.~\ref{fig:4}(b) and Appendix~\ref{app:B}]. Eq.~(\ref{eq:e_i}) is again equivalent to a power law in the spectral density, $\Omega_\act (K) \sim K^{-\beta^*}$, and is consistent with the affinity-rank statistics in randomly sampled individuals. The exponent $1/\beta^*$ equals the activation temperature of the naive repertoire, as given by Eq.~(\ref{eq:beta}). By combining Eqs. (\ref{eq:x_i}) and (\ref{eq:e_i}), we obtain a power-law relation between size and affinity, 

\BE
\langle \bar N_j\rangle \sim \langle K_j\rangle^{-\zeta\beta^*} 
\qquad \mbox{(HS)},  
\label{eq:N_K}
\EE
as shown in Fig.~\ref{fig:4}(c). In the HS regime, the size and affinity rankings coincide up to fluctuations, because both are related to time: high-affinity clones get activated before low-affinity clones. 
In the LS regime, the size-affinity correlation is lost. Large clones have affinities of order $K_p$, high-affinity clones have small size and show a faster decline of affinity with rank than in the HS regime [see Fig.~\ref{fig:3}(a), \ref{fig:4}(b) and~\ref{fig:4}(c)]. Hence, empirical observations of this correlation can provide specific evidence for activation by proofreading in the HS regime. 

\begin{figure*}[t]
\begin{center} 
\includegraphics[width= 1 \textwidth]{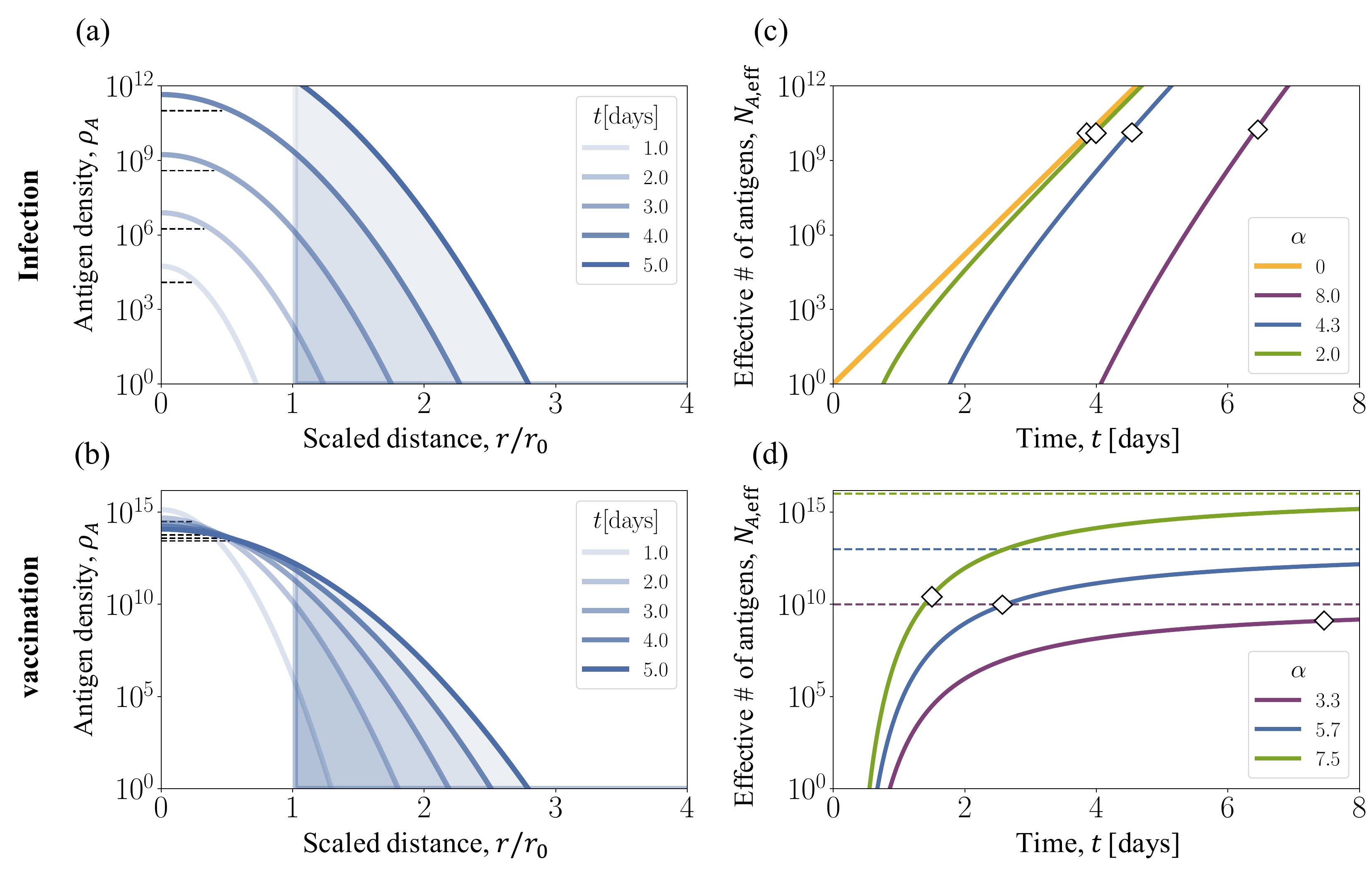}
\end{center}
\caption{ {\bf Spatio-temporal recognition dynamics.} 
{\bf (a, b)} Time-dependent antigen density $\rho({\bf r}, t)$, plotted as a function of the scaled distance from the antigen entry point, $r/r_0$. 
(a) Infection, 
(b)~immunization with an inactivated vaccine. Dashed lines indicate the diffusion range, $\sqrt{Dt}$; the shaded region contains the approximate number of antigen particles contributing to B cell activation, $N_{A, {\rm eff}} (t)$ (see text). 
{\bf (c, d)} Effective number of antigen particles, $N_{A, {\rm eff}} (t)$, for different values of the scaled recognition radius, $\alpha$. These functions are plotted together with the total antigen number, (c)~$N_A (t) = \lambda_A t$ (corresponding to the limit $\alpha = 0$, yellow line, same as in Fig.~\ref{fig:2}(a)) and (d)~$N_A (t) = N_0$ (dashed lines). Diamonds mark the start of B cell activation. 
Model parameters: 
(a, b) $r_0 = 0.5 {\rm cm}$; $D = 3\cdot 10^{-3} {\rm cm}^2{\rm s}^{-1}$; 
(c) $r_0 = 0.5 {\rm cm}$; 
$D = 6\cdot 10^{-4}, 3\cdot 10^{-3}, 1.5\cdot 10^{-2} {\rm cm}^2{\rm s}^{-1}$, 
corresponding to $\alpha = 2.0, 4.3, 8.0$; 
(d) $r_0 = 0.5 {\rm cm}$; $D = 3\cdot 10^{-3}{\rm cm}^2{\rm s}^{-1}$; $N_0 = 10^{10}, 10^{13}, 10^{16}$, 
corresponding to $\alpha = 3.3, 5.7, 7.5$.
Other parameters as in Fig.~2. 
\label{fig:5}
}
\end{figure*}

\vspace{1ex} \paragraph*{\bf Generalised Luria-Delbr\"uck processes.}
The antigen recognition statistics of a primary infection, as given by Eqs.~(\ref{eq:x_i}) to~(\ref{eq:N_K}), is characterised by two independent exponents, the size exponent $\zeta$ and the phenotype exponent $\beta_\act$. These statistics define a specific class of exponential processes, where growth is mediated by a recognition phenotype. We argue this class to be relevant for recognition of exponential signals, and refer to it as generalised Luria-Delbr\"uck processes. The analogy becomes clear by comparison with the proliferation-mutation statistics of a classical Luria-Delbr\"uck fluctuation experiment, where a wild-type cell population grows exponentially with rate~$\lambda_A$, cells mutate with a constant rate $U$, and mutant clones grow with rate $\lambda_B$. This process produces a power-law clone size statistics of the form of Eq.~(\ref{eq:x_i}) with size exponent
\BE
\zeta_0 = \frac{\lambda_B}{\lambda_A};  
\EE
[see Appendix~\ref{app:B} and ref.~\cite{Zheng1999}]. In the LS regime ($\beta_\act = p$), the clone size statistics of activated B cells follows the classical Luria-Delbr\"uck form, $\zeta = \zeta_0$. In the HS regime ($\beta_\act = \beta^*$), however, the B cell size exponent takes a different form, $\zeta = \zeta_0 \times (p/\beta^*)$, the correction factor reflecting the correlation between clone size and recognition phenotype (antigen affinity). The exponent $\beta_\act$, which governs the statistics of the recognition phenotype given by Eq.~(\ref{eq:e_i}), has no analogue in a classical Luria-Delbr\"uck process. 
This exponent enters the number of activated B cell clones, $L_\act (t)  \sim \exp [ (\beta_\act /p)\,\lam_A  t]$ [see Fig.~\ref{fig:2}(c)], which corresponds to the number of mutant clones in a classical Luria-Delbr\"uck process. Given a constant molecular clock of mutations, this number always grows with rate $\lam_A$, proportionally to the wild-type population size. 

\paragraph*{\bf Spatiotemporal antigen dynamics.}
A acute viral infection typically starts in narrowly localized spatial region. The spread of the antigen into the surrounding tissue is driven by proliferation and diffusion,  generating an increasing density, 
\BE
\rho_A (\x, t) = \frac{N_0}{(4 \pi Dt)^{3/2}} \, \exp \left (\lambda_A t - \frac{\x^2}{4 D t} \right ) ,
\label{eq:diff} 
\EE
where $N_0$ is the initial number of antigens, $D$ is the diffusion constant, and $\x$ the distance vector to the point of origination [see Fig.~\ref{fig:5}(a)]. Subsequently, antigen particles are drained in lymph vessels into the vicinity of B cells, most of which are located in secondary lymphoid organs like lymph-nodes and spleen~\cite{Pape2007}. This transport process also involves antigen capture by macrophages. Hence, antigen recognition requires a characteristic initial distance $r_0$, typically to a nearby lymph vessel, to be bridged by diffusion. Here we describe this constraint by a reduced, effective number of antigen particles interacting with B cells, $N_{A, {\rm eff}} (t) = \int_{|\x| > r_0} \rho_A (\x,t) \, {\rm d}\x$ [shaded region in Fig.~\ref{fig:5}(a)]. This number of antigens enters the activation dynamics described by Eqs.~(\ref{u_on})--(\ref{eq:P_act})  [see Fig.~\ref{fig:5}(b)]. The start of activation in the HS regime, described as before by the condition $R(K^*, t^*) \sim 1$, now takes place at a diffusion-dependent time, $t^* (D, r_0)$, with an effective growth rate, $\lambda^* (D, r_0) \equiv \dot N_{A, {\rm eff}} (t^*) / N_{A, {\rm eff}} (t^*)$. The statistics of activated B cell repertoires retains the form described above; the clone size exponent takes a diffusion-dependent value $\zeta (D, r_0)$ given by Eq.~(\ref{eq:zeta}) with the effective growth rate $\lambda^* (D, r_0)$ replacing the growth rate $\lambda_A$.

The spatio-temporal activation dynamics has two regimes, which are distinguished by the scaled recognition radius $\alpha = r_0/(D t^*)^{1/2}$. In the growth-dominated regime ($\alpha \lesssim \tilde \alpha$, typical activation events take place within the diffusive range, which implies a weak diffusion constraint on antigen recognition. In this regime, the activation start time $t^* (D, r_0)$, the effective growth rate $\lambda_A^* (D, r_0)$, and the resulting clone size exponent $\zeta (D, r_0)$ remain close to the values for the homogeneous system [Fig.~S4]. For $\alpha \gtrsim   \tilde \alpha$, activation requires rare diffusive paths, which causes a strong increase of $t^* (D, r_0)$ and $\lambda_A^* (D, r_0)$. A full analytical treatment of both regimes is given in Appendix~\ref{app:A}. For physiological parameters of the antigen dynamics in an acute infection ($N_0 \sim 1$, $D \sim 10^{-3} \mathrm{cm}^2 {\rm d}^{-1}$~\cite{Beauchemin2006}, and $r_0 \sim0.1- 1$ cm), the recognition dynamics is in the growth-dominated regime ($\alpha = 0.3 - 3.5  < \tilde \alpha \approx 10$) [see Fig.~S4]. Hence, the homogeneous-system calculus used above provides a good approximation for physiological antigen recognition processes in acute infections. 

\vspace{1ex} \paragraph*{\bf Immune response to vaccination.}
Following immunization with an inactivated vaccine, antigen spreads by diffusion but without proliferation [see Fig.~\ref{fig:5}(c)]. Here we analyze the antigen dynamics as a function of the vaccine dosage, $N_0$, 
with a diffusion constant similar to live antigens and near-zero growth ($\lambda_A \approx 0$, neglecting molecular decay on the relevant time scale). 
In vaccination-induced immune responses, 
the dosage and the diffusion parameters set the scaled recognition radius of antigen diffusion, $\alpha (D, r_0, N_0)$, the start of activation, $t^*(D, r_0, N_0) = (r_0^2/D) \, \alpha^{-2}$ and the effective growth rate, $\lambda^*(D, r_0, N_0) \simeq (D/r_0^2) \, \alpha^4$ ($\alpha \gtrsim 1$) [see Fig.~\ref{fig:5}(d), Fig.~S4 and Appendix~\ref{app:B}]. We conclude that the generalized Luria-Delbr\"uck model of activated repertoires also applies to vaccinations; a diffusion-dependent clone size exponent $\zeta (D, r_0, N_0)$ is now given by Eq.~(\ref{eq:zeta}) with the effective growth rate $\lambda^* (D, r_0, N_0)$.

Moreover, the spatio-temporal recognition model predicts a dosage window for successful vaccinations. For too low dosage ($\alpha \lesssim 1$, corresponding to $N_0 \lesssim (b_0 k_\on \rho_B)^{-1} (K^*/K_{\rm step})^p$), the vaccine fails to efficiently activate high-affinity B cell lineages because the onset of activation is delayed. For too high dosage ($N_0 \gg N_{\min}$), the activated repertoire gets strongly biased towards low-affinity clones, as indicated by small values of $\zeta$ [see Fig.~S4 and Appendix~\ref{app:A}]. For typical vaccine dosage values ($10\mu\mathrm{g}$, corresponding to $N_0 \sim 10^{13}$ particles~\cite{Tas2016, Mesin2020}), the recognition dynamics is in the diffusion-limited regime ($\alpha \approx 4 > \tilde \alpha \approx 1$; shaded region and dashed line in Fig.~S4 (c)-(f)). Below we apply this model to data of activated mouse repertoires induced by vaccination. 

\begin{figure*}[ht!]
\centering
\includegraphics[width=1\textwidth]{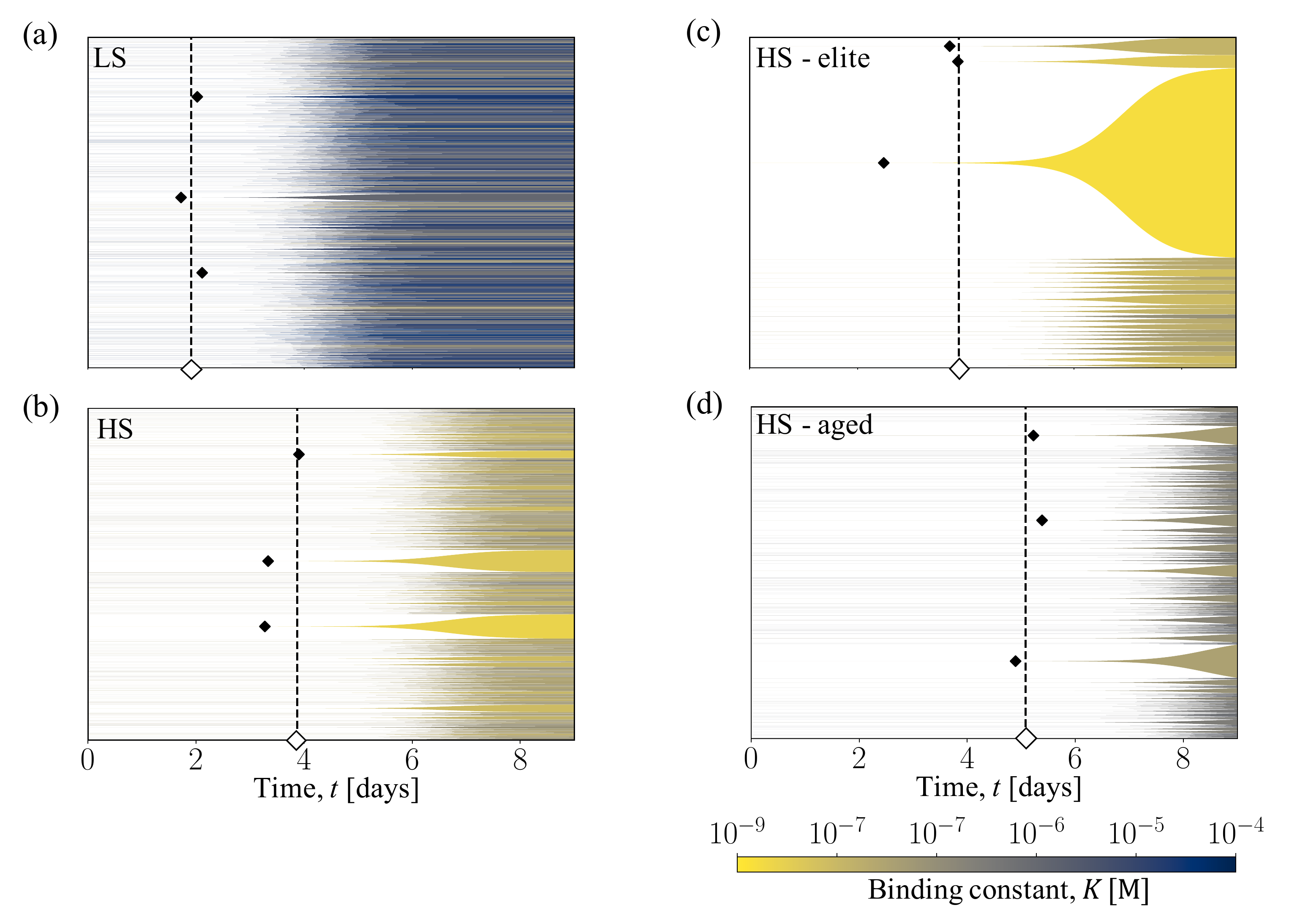}
\caption{ {\bf Activation patterns of B cell repertoires in an acute infection.}  
{\bf (a,b)}~Affinity and size of activated clones in randomly sampled individuals in the LS regime ($p = 1$) and in the HS regime ($p = 4$). Muller plots with filled areas representing individual clones (height: time-dependent population size, shading: antigen binding constant, $K$, as given by color bar); activation times of the first few clones ({filled diamonds}) are shown together with the expectation value $t_{\act}$ ({open diamond}). 
{\bf (c)} Activated clones in an elite neutraliser occurring at population frequency {$10^{-3}$}. This repertoire is marked by the early activation of a high-affinity ``jackpot'' clone.  
{\bf (d)} Activated clones of an aged individual with a 10x reduced number of naive lineages. The onset of activation, $t_\act$ ({black dashed line}), is delayed with respect to a full repertoire (panel b).  
Model parameters as in Fig.~\ref{fig:2}. 
}
\label{fig:6}
\end{figure*}

\section{Biological features of primary immune response}

\begin{figure*}[ht!]
\centering
\includegraphics[width=0.9\textwidth]{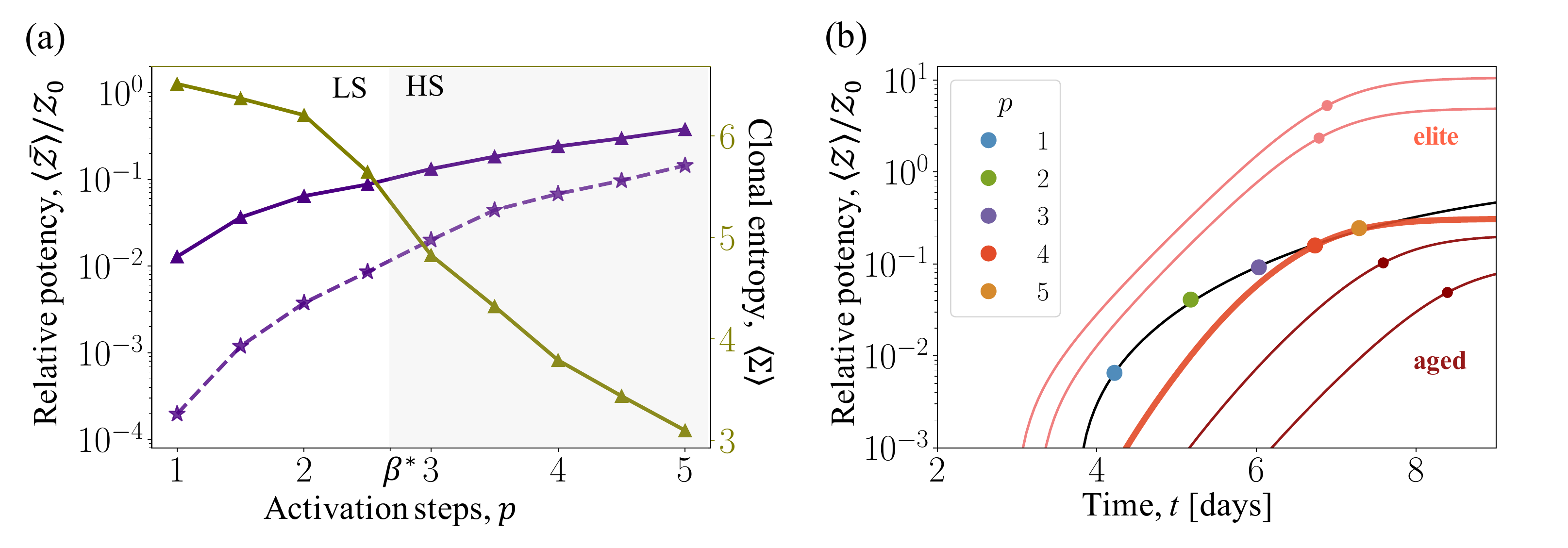}
\caption{ {\bf Proofreading determines potency, speed, and diversity of recognition.}  
{\bf (a)}~Antiserum potency, $\langle \Zbar \rangle/ K_0$, potency component of the largest clone, $\langle \zbar_1 \rangle$, and Shannon entropy of the activated repertoire, $\Sigma$. Population averages are shown as functions of the number of activation steps, $p$, at carrying capacity and relative to a reference value $\Z_0 = 1$M. Shading marks the HS regime ($p \gtrsim \beta^*$), an open circle with error bars gives the empirical entropy inferred from the repertoire data of ref.~\cite{Tas2016, Mesin2020}.
{\bf (b)}~Time-dependent antiserum potency, $\langle \Z \rangle (t)$. The population average is shown in the HS regime ($p = 4$, thick red line); the half-saturation point $(t_{50}, Z_{50})$ is marked by a dot. The family of half-saturation points for different values of $p$ characterises the tradeoff between potency and speed of typical immune responses, which defines a Pareto line (black line). The potency $\Z (t)$ of typical elite neutralisers (at population frequencies $10^{-3}$ and $10^{-4}$, light red lines) is above, that of aged immune systems (at 10x and 100x lineage reduction, dark red lines) is below the population average (HS regime, $p = 3$). Model parameters as in Fig.~\ref{fig:2}. 
}
\label{fig:7}
\end{figure*}

\vspace{1ex} \paragraph*{\bf Efficient immune response by kinetic proofreading.}  
In Figs.~\ref{fig:6}(a) and~\ref{fig:6}(b), we show simulation results for primary B cell activation in secondary lymphoid organs for typical individuals without proofreading ($p = 1$) and with proofreading at an intermediate number of steps ($p = 4$) [See Appendix~\ref{app:D} for details on the simulation procedure]. Shaded areas record the time-dependent size of activated B cell clones induced by an acute infection or vaccination. These clone dynamics are marked by an initially exponential growth and subsequent saturation given by the carrying capacity of the total activated repertoire~\cite{Bocharov1994}. Later stages of the antigen and B cell dynamics, including affinity maturation in germinal centers, antigen clearance, and subsequent degradation of activated B cells, are not relevant for the recognition dynamics discussed in this paper and not displayed here. 

Three effects of proofreading are immediately recognizable: it reduces the number of activated clones, increases the binding affinity of typical clones, and delays the onset of activation. These effects reflect the basic function of proofreading in the HS regime derived in the previous section:  deterministic activation of high-affinity lineages is coupled to strong suppression of low-affinity lineages. To quantify the impact of the activation dynamics on immune function, we evaluate the antiserum potency
\BE
\Z (t) = \sum_j \frac{N_j (t)}{K_j (t)} , 
\label{potency} 
\EE
as well as the contributions of individual lineages, $\z_j (t) = N_j (t) /K_j$ (the index $j = 1, 2, \dots$ orders clones by decreasing size). The function $\Z(t)$ sums the antigen affinities of all activated cells and determines the fraction of neutralised virions; it reaches a saturation value $\bar{\Z}$ at carrying capacity of the activated repertoire. We measure potency relative to a reference value $\Z_0$ describing a hypothetical repertoire with homogeneous binding constant $K^*$. 
In Fig.~\ref{fig:7}(a), we plot the population average $\langle \bar{\Z} \rangle$ as a function of the number of proofreading steps (here and below, population averages are denoted by brackets). Potency comes close to the reference value in the HS regime, but quickly drops with decreasing  $p$ in the LS regime. Without proofreading ($p = 1$), $\langle \bar{\Z} \rangle$ is about 20fold lower than at $p = 4$, argued below to be the approximate number of proofreading steps in human B cell activation. The difference between activation regimes is even more pronounced for the potency contribution of the largest clone, $\langle \bar{\z}_1 \rangle$ [see Fig.~\ref{fig:7}(a)]. In the HS regime, where the largest clone is likely also the clone of highest affinity, $\langle \bar{\z}_1 \rangle$ contributes a substantial fraction of the total potency; the HS potency-rank relation predicted by our model is shown in Fig.~S3. There is again a rapid drop in the LS regime; without proofreading, $\langle \bar{\z}_1 \rangle$ is about 1000fold lower than at $p = 4$. 

Another striking difference between the activation regime is in the activation probability of individual lineages, as given by the recognition function $R (K, t)$. In the HS regime, almost all high-affinity lineages get activated ($R$ reaches values close to 1); in the LS regime, most available high-affinity lineages do not get activated and are waisted for pathogen suppression ($R \ll 1$). Together, we conclude that the lineage activation profile generated by kinetic proofreading in the HS regime is a prerequisite for a potent, specific, and efficient primary immune response to an infection or vaccination.

\vspace{1ex} \paragraph*{\bf Repertoire-tuned proofreading.} 
Fig.~\ref{fig:6}(a) and~\ref{fig:6}(b) show two further functional differences between the activation regimes. In the LS regime, a large number of lineages gets activated, but these clones reach only small population frequencies at carrying capacity, $\bar x_j \equiv \bar N_j/\bar N$. In the HS regime, activation gets increasingly focused on few high-frequency and high-affinity lineages. We can describe the diversity of activated repertoires by the Shannon entropy $\Sigma = - \sum_j \bar x_j \log \bar x_j$. In the LS regime, the population average $\langle \Sigma \rangle$ is large and varies only weakly; in the HS regime, $\langle \Sigma \rangle$ drops substantially with increasing $p$ [see Fig.~\ref{fig:7}(a)]. Subsequent to activation, a part of the B cells undergoes affinity maturation in germinal centers. This mutation-selection process produces high-affinity plasma B cells, as well as a diverse set of memory B cells. In the HS regime, the larger repertoire diversity found close to the crossover point ($p = \beta^*$) serves both channels of affinity maturation: it facilitates the search for mutational paths towards high-affinity BCR genotypes in plasma cells, and it provides diverse input for memory cell formation~\cite{Shinnakasu2016}.

In Fig.~\ref{fig:7}(b) we plot the time-dependent, population-averaged potency for different values of $p$. The response time $t_{50}$, where $\langle \Z \rangle (t)$ reaches the half-saturation point $\Z_{50} = \langle \bar{\Z}/2 \rangle$, is marked by dots. The $p$-dependent increase in potency is coupled to an increased time delay of activation, caused by the sequence of intermediate steps and by the constraint to high-affinity lineages. In the HS regime, increasing $p$ yields a diminishing return of potency, while $t_{50}$ continues to increase proportionally to $p$. Similarly, efficient proofreading requires a sufficiently small activation rate. In the HS regime, for $k_\step < K^*k_\on$, decreasing $k_\step$ yields a diminishing return of potency, while $t_{50}$ continues to increase proportionally to $1/k_\step$ [see Fig.~S5]. The tradeoff between potency and speed of immune response defines a Pareto surface. This tradeoff, together with the entropy pattern, suggests that optimal immune response is tuned to the spectral density of the naive repertoire: the number and rate of proofreading steps are in the HS regime close to the crossover point $p = \beta^*$ and $k_\step = K^*k_\on$, respectively.

\vspace{1ex} \paragraph*{\bf Activated B cell repertoires in mice.}
While there is no direct evidence of proofreading in B cell activation to date, available data show activation patterns consistent with proofreading in the tuned regime. Specifically, we analyze recent data of clonal B cell populations in early germinal centres of mice after immunization~\cite{Tas2016, Mesin2020}. First, the clonal diversity of these populations can be characterized by a repertoire entropy $\Sigma =4.2\pm0.4$ [circle in Fig.~S6(a)]. Second, the empirical rank-size relation can be fit to a power law with exponent {$\zeta=0.50\pm 0.04$} [insert in Fig.~\ref{fig:4}(a)]. Strikingly, these data demonstrate that broad clone size distributions can be generated already in the first stages of a primary B cell immune response. 

Both of these summary statistics can be quantitatively explained by a generalized Luria-Delbr\"uck recognition dynamics with tuned proofreading. We recall from the previous section that the recognition dynamics induced by an inactivated vaccine involves an effective antigen growth rate at the locus of recognition, $\lambda^*(D, r_0, N_0)$, which depends on the antigen diffusion constant $D$, the effective distance $r_0$ to be bridged by diffusion, and the vaccine dosage $N_0$ [see also Appendix~\ref{app:B}]. Using empirical values of these parameters, we infer $\lam^*\approx 5.4 \, {\rm d}^{-1}$, similar to typical antigen growth rates in an infection. Together with physiological parameters $\lambda_B$ and $\beta^*$ characterizing the naive repertoire, the model predicts a proofreading-dependent repertoire entropy,  $\langle \Sigma \rangle (p)$, and a clone size exponent, $\zeta (p)$, by Eq.~(\ref{eq:zeta}) [Fig.~S6, to be compared with Fig.~4(d) and Fig.~7(a)]. The empirical values of $\Sigma$ and $\zeta$ obtained from the data of refs.~\cite{Tas2016, Mesin2020} are seen to match the model predictions for 
$p = 2.7 \, [2.2, 3.1]$, 
consistent with proofreading on the Pareto line of repertoire-tuned proofreading. Details of data analysis and inference of repertoire statistics are provided in Appendix~C.

Remarkably, clone size distributions extracted from data of human B cell repertoires~\cite{Briney2019}, which include memory-induced clone activation, show power laws with a similar exponent, {$\zeta=0.57\pm0.12$}~\cite{Chardes2022}.  Given that memory cells are in the same affinity range than activated naive cells~\cite{Mesin2020, Viant2020}, this may point to a common dynamical mechanism generating power laws in early immune responses. However, our present model contains only the primary activation dynamics and is not directly applicable to memory cells. In refs.~\cite{Desponds2016, Chardes2022}, the power laws of memory repertoires have been attributed to long-term selection, which requires multiple recurrent infections affecting the same set of B cell lineages.

\vspace{1ex} \paragraph*{\bf Elite neutralisers.}
Generalised Luria-Delbr\"uck immune activation shows particularly pronounced variation between hosts. In the HS regime, a subset of {\em elite neutralisers} is distinguished by early activation of a single high-affinity clone. This jackpot clone dominates the activated immune repertoire and generates exceptionally high potency. Fig.~\ref{fig:6}(c) shows an example of the activation dynamics that occurs in one of $10^{3}$ individuals, which is to be compared with the pattern in typical individuals [see Fig.~\ref{fig:6}(b)].  
Such elite neutralisers are ahead of the Pareto surface of typical immune responses [see Fig.~\ref{fig:7}(b)]. The cumulative distribution $\Phi (\Z)$, which gives the fraction of responders with saturation potency $> \Z$, displays two regimes of elite neutralisers [see Fig.~S3]. In the pre-asymptotic regime, the jackpot clone takes a large part but not all of the repertoire ($\bar N_1 < \bar N$). The pre-asymptotic potency distribution turns out to be dominated by clone size fluctuations, which implies {$ \Phi (\Z) \sim \Z^{-1/\zeta}$ }[see Appendix~\ref{app:B}]. In the asymptotic regime, the jackpot clone dominates the repertoire ($\bar N_1 \simeq \bar N$); hence, $\Phi (\Z)$ is proportional to the naive density of high-affinity clones, 
$\Phi (\Z) \sim L_0 (\bar N / \Z)$. For example, one in $10^5$ individuals has a primary response with potency 100x above average, comparable to a memory immune response carried by affinity maturated B cell lineages. 

\vspace{1ex} \paragraph*{\bf Immune ageing.} 
Recent results indicate that the most prominent effect of immune ageing is a decrease in the overall size and diversity of the repertoire~\cite{Gibson2009, Weiskopf2009, Wang2014, Bradshaw2022}. Our model predicts two effects of this decrease: primary immune responses come later and with reduced potency. Simulations of the activation dynamics in an aged repertoire in the HS regime show that the activation is delayed and antigen affinities are reduced compared to a full-size repertoire [see Fig.~\ref{fig:6}(d), to be compared with Fig.~\ref{fig:6}(b)]. The time-dependent potency remains behind the Pareto line and reaches a reduced value at carrying capacity [see Fig.~\ref{fig:7}(b)]. For a 10fold decrease in repertoire size, $t_{50}$ increases by about 1d and $\langle \bar{\Z} \rangle$ drops to half of its full-size value. Fig.~S5 shows the full dependence of potency on repertoire size. Given a reference point in the HS regime, a reduction of size always has a sizeable effect, while an increase eventually induces a cross-over to the LS regime and a diminishing return of potency.

\section{Discussion}

A potent adaptive immune response to a new antigen requires the specific activation of immune cells with high affinity to the antigen. Here we have developed a minimal, biophysical model for immune recognition by naive B cells in a primary infection or vaccination [see Fig.~\ref{fig:1}]. We have shown that active processes of antigen recognition -- kinetic proofreading -- are essential to constrain a primary immune response to high-affinity lineages. A possible proofreading mechanism is macrophage-mediated active transport of multiple antigen particles to the same B cell. The T cell response of an acute infection is expected to follow a similar kind of exponential signal recognition dynamics, albeit at overall lower affinities and by different molecular pathways. This will be the subject of a future publication.

A highly specific real-time response to a new antigen requires proofreading with depth above a threshold value, $p > \beta^*$.  Available data of mouse B cell repertoire activation are consistent with proofreading close to the specificity threshold, $p \approx \beta^*$, which amounts to about 3 consecutive proofreading steps [see Figs.~\ref{fig:4} and \ref{fig:7}]. As shown by our model, immune responses tuned to this point are close to a functional optimum: they balance potency and speed, and they generate a diverse set of activated clones for subsequent affinity maturation.

Our model treats the proliferation-activation dynamics of a primary immune response as a generalised Luria-Delbr\"uck process [see Fig.~\ref{fig:2}]. This class of processes is argued to be relevant for the molecular recognition of exponential signals. It is characterised by fluctuations coupled to recognition function: jackpot clones have large size and high affinity to the recognition target, here the exponentially growing antigen population [see Fig.~\ref{fig:6}(c) and~\ref{fig:7}(b)]. The underlying statistics of activated immune repertoires is characterised by statistical power laws with two basic exponents: 
the size exponent $\zeta$ and the phenotype exponent $\beta^*$. 
These exponents characterise the proofreading-dependent specificity of activated immune repertoires, as given by Eqs.~(\ref{eq:x_i}) and~(\ref{eq:e_i}). Importantly, our model predicts $\zeta$ and $\beta^*$ in terms of independently measurable quantities and without fit parameters. The statistics of activated B cell repertoires shape clinically important characteristics of primary infections and vaccinations, including the potency drop of aged responders and the increased response of elite neutralisers [see Fig.~\ref{fig:7}]. 

Power-law statistics of lineages and the occurrence of elite neutralisers are commonly ascribed to antigen-mediated selection on immune repertoires, often through multiple exposures~\cite{Doria-Rose2014, Desponds2016, Nourmohammad2016, Yan2020, Chardes2022}. Here we have shown that similar features can already emerge in primary immune responses to an acute infection or vaccination, prior to any antigen-mediated selection effects. Following the primary response, a part of the activated B cell repertoire is further processed by affinity maturation. This step is again driven by non-equilibrium antigen recognition processes~\cite{Jiang2023}.

Repertoire sequencing combined with neutralisation assays can test our model and probe adaptive immune systems in new ways. By recording the power-law rank-size relation of linages in early post-infection B cell repertoires and measuring the antigen binding constant of these clones, we can extract the corresponding power laws and infer the central parameters of our model: the specificity threshold, $\beta^*$ and the number of proofreading steps, $p$ [see Fig.~\ref{fig:4}]. The parameter $\beta^*$ measures the density of B cell lineages close to the maximum-affinity lineage. This density is set by the {\em global architecture of B cell immunity:}  the size of the naive repertoire and the complexity of the antigen-receptor binding motif. Both quantities emerge as key determinants of primary immune responses. In contrast, the parameter $p$ characterises the {\em molecular dynamics of antigen recognition}. As we have shown, $p$-step proofreading generates an effective inverse temperature $\beta \approx p$ that measures the specificity gain in a dense repertoire. At the point of optimal recognition, $p \sim \beta^*$, antigen recognition dynamics matches repertoire complexity. From this point, increasing $p$ at a constant repertoire size $L$ produces a diminishing return of proofreading; conversely, increasing $L$ at constant $p$ produces a diminishing return of repertoire size. 

The infection response dynamics described in this paper is a direct target of the co-evolution between pathogens and host immune systems. For respiratory viruses, infection characteristics relevant for pathogen fitness include the duration and viral load of the symptomatic infectious period~\cite{Park2019, Park2022}. Given the speed-specificity tradeoff of host response [see Fig.~7], viruses can increase fitness by increasing the within-host reproduction rate $\lam_A$. The clearance of the virus involves antibody binding; the resulting fitness effects depend on the distribution of antigen binding affinities, $K$, in the host population, which in turn depends on age-dependent repertoire sizes. The statistics of activated repertoires developed here serves to stratify antigenic interactions in populations structured by age and infection history, setting the molecular basis for escape evolution from immune recognition.

Finally, the link between repertoire size and antigen recognition machinery has implications for the macro-evolution of adaptive immune systems.

The size of the total B cell repertoire varies drastically across vertebrates, ranging from $\sim3 \times 10^{5}$ cells in zebrafish~\cite{Weinstein2009} to $\sim 10^{11}$ in humans~\cite{Altan-Bonnet2020, Glanville2009, Elhanati2015}. Because potency and timeliness of immune responses are likely to be under strong selection, the functional balance of tuned proofreading, $p \approx \beta^*$, is also expected to be a maximum of fitness. Evolutionary changes of the repertoire size are then predicted to occur together with changes of the recognition machinery, which includes number and rate of proofreading steps. Tracing these co-evolutionary dynamics by comparative cross-species studies may provide a new avenue to understand the evolution of complex immune systems. 

\begin{acknowledgments}
We thank R. Andino, A. Ryabova, and D. Valenzano for discussions, M.~Karmakar and M.~Meijers for a careful reading of the manuscript, and all members of the L\"assig lab for input. This work was partially supported by Deutsche Forschungsgemeinschaft (grant CRC 1310, to ML).
\end{acknowledgments}

\appendix

\section{Antigen recognition dynamics} 
\label{app:A}

\vspace{1ex} \paragraph*{\bf Antigen-receptor interaction models.} 
Here we model the binding (free) energy between an antigen with epitope sequence $\a = (a_1, \dots, a_\ell)$ and a B cell receptor (BCR) sequence $\b = (b_1, \dots, b_\ell)$ as an additive function, 
\BE
\Delta E (\a, \b) = \sum_{k = 1}^\ell \varepsilon (a_k, b_k). 
\label{E} 
\EE
This function includes entropy contributions of non-translational degrees of freedom; i.e., rotations and elastic deformations of the molecules involved. We use three established models of amino acid interactions $\ep (a, b)$: the TCRec model originally inferred for T cell receptors~\cite{Karnaukhov2022}, the Miyazawa-Jernigan matrix~\cite{Miyazawa1996}, and normally distributed random energies. In each case, we introduce a scale factor that is inferred from measured BCR-antigen binding energies (see below). The zero point of $\Delta E$, by definition, corresponds to a reference antigen-BCR pair with equilibrium binding constant $K_0 = 1$M. In this gauge, the binding energy and the dissociation constant $K$ of an arbitrary pair are related by 
\BE
\frac{K(\a, \b)}{K_0} = \exp \left [\frac{\Delta E (\a, \b)}{k_BT} \right ], 
\label{EK} 
\EE
where $k_B$ is Boltzmann's constant. In the main text and below, we express all energies in units of $k_BT$ at a fixed, physiological temperature $T$.

\vspace{1ex} \paragraph*{\bf Activation of B cells by kinetic proofreading.} 
An early infection is characterised by low densities of antigens and B cells. Accordingly, we model the activation of individual B cells upon functional binding with a single antigen. We assume that activation of antigen-bound cells requires a chain of $p$ irreversible steps with characteristic rate $k_\step$ [see Fig.~\ref{fig:1}]. Hence, the B cell activation rate takes the form 
\BE
u_\act(K, t) = u_\on(t) \, p_a(K). 
\label{eq:u_act}
\EE
Here, the association rate per B cell lineage, 
\BE
u_\on(t) = N_A(t)b_0 k_\on \rho_B, 
\label{u_on2} 
\EE
is proportional to the number of antigen particles, $N_A$, the number of receptors per cell, $b_0$, the diffusion-limited association rate to a single B cell receptor, $k_\on$, and the lineage specific B cell density, $\rho_B$. The probability of activation after association is given by 
\BE
p_a = \left(\frac{k_\step}{k_\step+k_\off} \right )^{p} = \left(\frac{1}{1+K/K_\step}\right)^{p}, 
\label{eq:p_a}
\EE
where $k_\off=K k_\on$ and $k_\step=K_\step k_\on$. At each intermediate state, the antigen can dissociate or undergo the next activation step; these alternatives are independent Poisson processes with rates $k_\off$ and $k_\step$, respectively. Hence, the next activation step occurs before dissociation with probability $k_\step /(k_\step+k_\off)$.

The probability that a lineage gets activated up to time $t$ is 
\BEA
R(K, t) &=& 1-  \exp{\left [- \int_0^t   u_\act(K, t') \, \d t' \right ] }\nonumber\\
&=& 1-  \exp{\left[-\frac{u_\act(K, t)}{\lam_A}\right]}, 
\label{eq:recognition}
\EEA
as given by Eq.~(\ref{eq:P_act}) of the main text. Here we have used that antigens proliferate exponentially, $\rho_A (t) \sim \exp(\lambda_A t)$. At early times, activation is association-limited and rare for all $K$, 
\BEA
R(K, t) & \simeq & N_A (t) \frac{b_0 k_\on \rho_B}{\lam_A(1+K/K_\step)^p} \label{eq:R1}
\\
& < & \exp \left [- \lam_A(t_1 - t) \right ], 
\qquad (t < t_1)
\EEA
where 
\BE
t_1 = \frac{1}{\lambda_A} \log \left ( \frac{\lam_A}{b_0 k_\on \rho_B} \right ). 
\label{t1}
\EE
For $t > t_1$, we have 
\BE
 R(K, t) \simeq \left \{ 
 \begin{array}{ll}
1 &( t > t_1, K < K_1(t)) 
\\ \\
\left ( \dfrac{K}{K_1(t)} \right )^{-p}  & (t > t_1, K > K_1 (t)),
\end{array} 
\right .
\label{eq:R2}
\EE
with 
\BE
K_1 (t)= K_\step \exp \left [ \frac{\lam_A}{p} \, (t - t_1) \right ]. 
\label{eq:K_R}
\EE
That is, deterministic activation of B cell lineages occurs along a moving front, $K_1 (t)$. Ahead of the front, activation is strongly suppressed by proofreading.

\vspace{1ex} \paragraph*{\bf Alternative models of B cell activation. } 
To highlight the specific role of kinetic proofreading in the activation of naive B cells, we compare the proposed activation mechanism to alternative mechanisms with reversible binding kinetics [see Fig.~S1]. The corresponding rates govern transitions between unbound and intermediate antigen-bound B cell states [marked by grey shading in Fig.~\ref{fig:1} and Fig.~S1]. We note that all activation mechanisms have at least one irreversible step: the last transition to exponential proliferation and antibody production (marked by green shading). 

In an early primary infection, the activation of B cells takes place under specific physiological conditions: 
(i)~The antigen density and, hence, the equilibrium occupancy of B cells remains low ($\rho_A/K \lesssim 10^{-1}$, given $\rho_A\approx 10^{11} / {\rm l} \lesssim 10^{-13}$M and $K \gtrsim K^*\approx 10^{-7}$M). These conditions differ drastically from the densities in confined spaces, e.g., lymph nodes and germinal centers, which are relevant for the binding kinetics of presented antigens. 
(ii)~The association kinetics of virions and plasma B cells is believed to be diffusion-limited; i.e., it takes place at a homogeneous rate $k_\on$ \cite{Hearty2012}. This excludes mechanisms for specific recognition by formation of immunological synapses, which have been proposed for presented antigen and operate by modulation of an activation-limited rate $k_\on$~\cite{Qi2001, Batista2001, Carrasco2004, Goldstein2004, Fleire2006}.
(iii)~For efficient activation, the rate $k_\step$ cannot be smaller than all other transition rates, as it is usually assumed in models of kinetic proofreading~\cite{Hopfield1974, Ninio1975}. Tuned rates discussed in the main text are of order $k_\step = 4\cdot 10^{-3}\text{s}^{-1}$, which implies $k_\step \gtrsim k_\off$ for high-affinity B cell lineages. In this regime, the activation dynamics at a given antigen density $\rho_A$ is close to a non-equilibrium steady state, even if the binding kinetics satisfies detailed balance. We consider two specific classes of models with diffusion-limited association and reversible binding kinetics: 

\paragraph*{Activation via an excited intermediate state.} This process is a reversible analogue of the proofreading dynamics discussed in the main text. Allowed transitions are between the unbound state and the primary bound state (with rates $k_\on, k_\off$) and between the primary bound state and the excited intermediate state (with rates $k_e^+, k_e^-$), as shown in Fig.~S1. Using detailed balance, the antigen-bound states have reduced binding energies $\Delta E = \log (K/K_0)$ and $\Delta E_e  = \Delta E + \Delta \Delta E_e$ respectively, where $K = k_\off / k_\on$ and $\Delta \Delta E_e = \log (k_e^-/ k_e^+)$. Like the proofreading model, this reversible model has a deterministic activation front given by Eqs. (\ref{t1}) and (\ref{eq:K_R}). However, the activation probability is asymptotically proportional to the equilibrium occupancy of the intermediate state, 

\BEA
R (K, t) &\simeq& N_A (t) \, \frac{b_0 k_\step}{\lambda_A} \, \frac{\rho_B}{K_0} \, \exp \left (- \Delta E_e \right )\nonumber\\
&\sim& \frac{N_A(t) \rho_B}{K} 
\qquad (K \gg K_1 (t)), 
\EEA
leading to weak suppression of low-affinity lineages~\cite{Hopfield1974, Ninio1975}.

\paragraph*{Activation by cooperative binding.} In this process, activation requires binding of two or more virions to receptors of the same B cell, which has been observed for antigens actively transported to lymph nodes and presented to B cells~\cite{Batista2001, Carrasco2004, Gonzalez2011}. A bound state of $p$ virions has the reduced binding energy $\Delta E_p = p \Delta E + \Delta \Delta E_p = \log (K_p /K_0)$, where $\Delta E = \log (K/K_0)$ is the single-particle binding energy and $ \Delta \Delta E_p$ is the contribution of cooperative binding. Fig.~S1 shows the case $p = 2$, where $\Delta \Delta E_p = \log(k'_\off / k_\off)$. In the cooperative binding model with detailed balance, the asymptotic activation probability is proportional to the equilibrium occupancy of the $p$-virion bound state, 

\BEA
R (K, t) &\simeq&  N_A (t) \,  \frac{b_0 k_\step}{\lambda_A} \, \frac{\rho^{p-1}_A (t) \rho_B}{K_0^p} \, \exp \left (- \Delta E_p \right )\nonumber\\
&\sim&  \frac{\rho^p_A(t)}{K^p} 
\qquad (K_p \gg K_\step),
\EEA
with $K_\step = k_\step / k_\on$. For $p > 1$, this model leads to stronger suppression of low-affinity lineages; however, at the low antigen concentrations of an early infection, even high-affinity naive lineages do not reach deterministic activation ($R \ll 1$ for $\rho_A (t)/ K \lesssim 10^{-1}$). 

We conclude that the kinetic proofreading mechanism of B cell activation introduced in the main text is the simplest model to generate deterministic activation of high-affinity lineages together with strong suppression of low-affinity lineages under the physiological conditions of an early primary infection. 

\vspace{1ex}\paragraph*{\bf Spatiotemporal recognition dynamics.}
The minimal spatio-temporal antigen dynamics introduced in the main text captures two key inhomogeneities relevant for immune recognition: acute infections and vaccinations start with a narrowly localized antigen distribution and B cell recognition takes place predominantly in secondary lymphoid organs. The minimal model describes the initial diffusion and proliferation of antigen particles through tissue, starting from an initial particle number $N_0$. This process generates a time-dependent antigen density $\rho_A (\x, t)$ given by 
\BE
\frac{\partial}{\partial t} \, \rho_A(\x, t) = D\nabla^2 \rho_A(\x, t) + \lam_A \rho_A(\x, t), 
\label{eq:reac-diff}
\EE
where $D$ is the diffusion constant. 
The solution of this equation is given by Eq.~(\ref{eq:diff}). Diffusion takes place over a characteristic distance $r_0$ from the starting point of the immunization to a nearby lymph vessel. Our model uses a simple approximation for the diffusive constraint on recognition: antigen particles are counted as interacting with B cells when they have reached a diffusive displacement $|{\bf r} | > r_0$. This condition defines an effective antigen number $N_A (t; D, r_0)$. We neglect details of the subsequent drainage dynamics in lymph vessels, which can be absorbed into a lineage-independent effective value of the association rate $k_\on$. A key quantity of the model is the scaled recognition radius, 
\BE
\alpha (D, r_0) = \frac{r_0}{\sqrt{D t^* (D, r_0)}}, 
\label{eq:alpha}
\EE
which is defined as the ratio of $r_0$ and the diffusion range at the start of activation. This parameter delineates two regimes of recognition: weakly diffusion-limited ($\alpha \lesssim 1$) and strongly diffusion-limited ($\alpha \gg 1$). It determines the effective number of antigen particles interacting with B cells at the start of activation, 
\BE
N_{A, {\rm eff}} (t^*; D, r_0) = \frac{N_A (t^*)}{ \sqrt{4\pi}} \alpha^3 I_0(\alpha), 
\label{eq:N_Aeff}
\EE
where $N_ (t)$ is the total number of antigen particles and 
\BE
I_\kappa(\alpha) \equiv \int_1^\infty e^{-x^2\alpha^2/2}x^{2+\kappa}\d x. 
\EE
In analogy to Eqs.~(\ref{u_on2}) to~(\ref{eq:recognition}), the onset of activation in the HS regime is determined by the condition $R(t^*, K^*) \sim 1$, 
here evaluated as 
\BE
\frac{u_\act (t^*, K^*)}{\lambda^*} = \frac{N_{A, {\rm eff}} (t^*; D, r_0)}{N^*_{\rm inf}}  = 1
\label{onset_st} 
\EE
with 
\BE
N^*_{\rm inf} = \frac{\lambda^*}{b_0 k_\on \rho_B} \left (1 + \frac{K^*}{K_{\rm step}} \right )^p. 
\label{Nstar} 
\EE 
This condition determines the onset time of activation, $t^* (D, r_0)$, and the corresponding effective antigen growth rate, $\lambda^* (D, r_0) \equiv \dot N_{A, {\rm eff}} (t^*; D, r_0) / N_{A, {\rm eff}} (t^*; D, r_0)$, for the spatio-temporal process. 

In the case of infections with antigen of initial particle number $N_0 = 1$ and proliferation rate $\lambda_A$, we obtain a closed solution by using the approximation $\lambda^* \approx \lambda_A$ in the onset condition (\ref{onset_st}). The onset time of activation then takes the form 
\BE
t^* (D, r_0) = t^* \, g_{t, {\rm inf}} (\alpha)
\label{eq:tstar}
\EE
where $t^*$ is the onset time in the homogeneous system given by Eq.~(\ref{eq:act2}) and 
\begin{eqnarray} 
g_{t, {\rm inf}} (\alpha) & = &  1-\frac{1}{\log N^*_{\rm inf}}\log\left[\frac{1}{\sqrt{4\pi}}\alpha^3I_0(\alpha)\right]
\nonumber \\
& \simeq & \left \{ 
\begin{array}{ll}
1 + O(\alpha^{3}) & (\alpha \lesssim \tilde \alpha),  
\label{tstar_inf} 
\\
\frac{1}{4\log N^*_{\rm inf}} \alpha^{2} &  (\alpha \gtrsim \tilde \alpha). 
\end{array}
\right.
\end{eqnarray} 
Similarly, the effective growth rate at the onset of activation is given by 
\BE
\lambda^* (D, r_0) = \lambda_A \, g_{\lambda, {\rm inf}} (\alpha)
\label{eq:lambdastar}
\EE
with 
\begin{eqnarray} 
g_{\lambda, {\rm inf}} (\alpha) & = & 1+ \frac{\alpha^2}{\alpha^2 + 4\log{\left [N^*_{\rm inf} \alpha^2 \sqrt{4\pi} \right]}} 
\nonumber \\
& \simeq & \left \{ 
\begin{array}{ll}
1  & (\alpha \lesssim \tilde \alpha),  
\\
2 &  (\alpha \gtrsim  \tilde \alpha).
\end{array}
\right.
\label{lambdastar_inf} 
\end{eqnarray} 
In (\ref{tstar_inf}) and (\ref{lambdastar_inf}), the crossover scale $\tilde \alpha$ is given by the condition 
\BE
\tilde \alpha^2 = 4\log \left[\sqrt{4\pi} N^*_{\rm inf} \tilde \alpha^2 \right ].
\label{alpha_tilde} 
\EE
The solution (\ref{eq:tstar}) -- (\ref{alpha_tilde}) is plotted in Fig.~S4(a, b) together with numerical evaluations of the onset condition (\ref{onset_st}). As expected, the onset time and the effective growth rate increase with increasing diffusive constraint. However, $\lambda^* (D, r_0)$ remains always close to $\lambda_A$, justifying the above approximation for the condition (\ref{onset_st}). 

In the case of vaccinations with antigen of initial particle number $N_0 \gg 1$ and proliferation rate $\lambda_A = 0$, rapid antigen growth at the locus of recognition emerges in the diffusion-limited regime ($\alpha \gtrsim 1$). In this regime, we can again write the time and the effective growth rate in scaling form, 
\BE
t^* (D, r_0, N_0) = \frac{r_0^2}{D} \, \alpha^{-2}
\qquad (\alpha \gtrsim 1)
\label{eq:t*vac}
\EE
the corresponding effective growth rate is given by 
\BE
\lambda^* (D, r_0,N_0) = \frac{D}{r_0^2} \, g_{\lambda, {\rm vac}} (\alpha)
\qquad (\alpha \gtrsim 1)
\label{eq:lambdastar_vac}
\EE
with 
\begin{eqnarray} 
g_{\lambda, {\rm vac}} (\alpha) & = &\frac{\alpha^4}{4}\left(\frac{I_2(\alpha)}{I_0(\alpha)} -\frac{6}{\alpha^2}\right)
\nonumber \\
& \simeq & \frac{\alpha^4}{4}  
\qquad  (\alpha \gtrsim  1). 
\label{lambdastar_vac2} 
\end{eqnarray} 
In this case, the onset time decreases, while the effective growth rate increases with increasing recognition radius; see Fig.~S4(c, d). The effective recognition radius now also depends on the vaccine dosage. From Eqs.~(\ref{onset_st}), (\ref{Nstar}), (\ref{eq:lambdastar_vac}, and (\ref{lambdastar_vac2}), we obtain the relation 

\BE
\alpha^2  \approx  4\log{\left[\frac{N_0}{N^*_{\rm vac}} \, \alpha^{-2} \right]}
\qquad (\alpha \gtrsim 1).
\label{alpha_dosage} 
\EE
with 
\BE
N^*_{\rm vac} = \frac{\sqrt{4\pi} D}{4 r_0^2 b_0 k_\on \rho_B} \left(1 + \frac{K^*}{K_\step}\right)^{p} . 
\EE
In Fig.~S4(e, f), we plot the resulting dosage-dependence of the activation time $t^* (D, r_0, N_0)$ and the effective growth rate $\lambda^* (D, r_0, N_0)$. A dosage window for successful vaccination emerges: for too small dosage ($\alpha \lesssim 1$, corresponding to $N_0 \lesssim N^*_{\rm vac}$), the onset time $t^*(D, r_0, N_0)$ becomes large; for too large dosage ($N_0 \gg N^*_{\rm vac}$), the clone size exponent $\zeta (D, r_0, N_0)$ becomes small, distorting the spectrum of activated lineages towards weak antigen affinity.

\section{Antigen recognition statistics} 
\label{app:B}

\vspace{1ex} \paragraph*{\bf Density of BCR states.}
To characterise the naive B cell repertoire available for response to a given antigen $\a$, we use the density of lineage states defined by a unique BCR sequence $\b$,
\BE
\Omega_0 (\Delta E) = \frac{\d}{\d \Delta E} \, L_0 (\Delta E),
\EE
where $L_0 (\Delta E)$ is the expected number of lineages in an individual with binding constant $K < \exp (\Delta E)$ to the epitope $\a$ (we suppress the dependence on $\a$ in this paragraph). By Eq.~(\ref{EK}), this form is equivalent to the definition given in the main text, 
$\Omega_0 (K) = (K \d / \d K) L_0 (K)$. 
We further define the micro-canonical entropy 
\BE
S(\Delta E) = \log \Omega_0 (K)
\label{SE} 
\EE
and the associated micro-canonical inverse temperature,
\BE
\beta (\Delta E) = \frac{\d S(\Delta E)}{\d \Delta E}, 
\label{betaE} 
\EE
a parameter that is independent of the physiological temperature $T$ appearing in Eq.~(\ref{EK}). Because we measure energies in units of $k_B T$, the parameter $\beta$ gives the inverse temperature in units of $(k_BT)^{-1}$. To compute these micro-canonical quantities, we evaluate the canonical partition function,
\BEA
Z (\beta) &=& \sum_\b \exp[- \beta \Delta E(\a, \b)]\nonumber\\
&=& \prod_{i = 1}^\ell \sum_b \exp[- \beta \varepsilon (a_i, b)],  
\label{Z} 
\EEA
which depends on $\beta$ as an independent parameter. This function defines the canonical binding energy,
\BE
\Delta E_c (\beta) = - \frac{\d}{\d \beta} \log Z(\beta), 
\label{Ebeta} 
\EE
which is an expectation value in the ensemble (\ref{Z}), and the associated entropy, 
\BE
S_c (\beta) = \log Z (\beta) + \beta \Delta E_c (\beta). 
\label{Sbeta} 
\EE
We invert the relation (\ref{Ebeta}) to write the inverse temperature as a function of the binding energy, $\beta (\Delta E_c)$, and we substitute this function into Eq.~(\ref{Sbeta}) to obtain $S_c (\Delta E_c)$. Upon equating $\Delta E = \Delta E_c$, these functions provide an excellent approximation to their micro-canonical counterparts $\beta (\Delta E)$ and $S(\Delta E)$, as given by Eqs. (\ref{SE}) and (\ref{betaE}). Thus, the canonical formalism provides an efficient way to compute the density of states, $\Omega_0 (\Delta E)$,  for the system at hand.

\vspace{1ex} \paragraph*{\bf Antigen-receptor ensembles.} 
To compare the response repertoires for different antigens, we evaluate the BCR lineage density $\Omega_0 (K, \a)$ for a random sample of epitope sequences $\a$ in a repertoire of overall size $L_0$. For a given amino acid interaction matrix and a given antigen, our energy model has two free parameters, the binding length $\ell$ and the scale factor of the energy, which sets the energy variance 
$\sigma^2_\varepsilon = \sum_k [ \sum_{b} \varepsilon^2 (a_k, b)/20 -  (\sum_{b} \varepsilon (a_k, b)/20)^2]$. Here we calibrate these parameters by tuning the minimum binding constant expected in an individual repertoire and the global minimum binding constant to observed values of typical high-affinity antibodies generated in primary infections and of ultra-potent antibodies, $K^* \approx 10^{-7}$M and $K_m \approx 10^{-11}$M~\cite{Eisen2014, Altan-Bonnet2020}.

The resulting ensemble of response repertoires has the following properties [see Fig.~S2]: 
(i) The distributions of inferred binding lengths and of the rms. energy variation per site are strongly peaked around values $\ell \sim 20$ and $\sigma_\epsilon / \ell^{1/2} \sim 1$. 
(ii)~A higher energy variance per site can be traded for a shorter binding length, consistent with a constraint on the total energy variance $\sigma^2_\varepsilon$. 
(iii)~The lineage densities $\rho_0 (K)$ depend only weakly on the antigen sequence $\a$ and have similar inverse activation temperatures, $\beta^* \equiv \beta (K^*) = 2.5 \pm 0.3$ for $L_0 = 10^9$. 

\vspace{1ex} \paragraph*{\bf Activation statistics of B cell repertoires.} 

Given the density of naive B cell lineages, $\Omega_0 (K)$, and the activation probability $R(t, K)$, we can evaluate the density of activated lineages, 
\BE
\Omega_\act(K, t)= \Omega_0(K) \, R(K, t),
\EE
and the total number of activated lineages, $L_\act  (t) = \int \Omega_\act (K,t) \, \d(\log K)$. Two activation regimes emerge:

\paragraph*{Low-specificity (LS) regime.}
In this regime, the specificity of activation is limited by the number of proofreading steps. According to Eq.~(\ref{eq:R1}), the function $\Omega_\act$ is strongly peaked at a value $K_p$ defined by the condition $\beta(K_p) = p$ [see Fig.~\ref{fig:3}(a)]. Integrating this function yields the expected number of activated lineages at time $t$,  
\BEA
L_\act(t) &=& \int \Omega_\act (K, t) \, \d(\log K) \quad \mbox{(LS})\nonumber\\
& \approx &  \frac{b_0 k_\on \exp{[\lam_A t]}}{{\cal N} }\int\frac{\Omega_0(K)\, \d(\log K)}{(1+K/K_\step)^p} \nonumber\\
&=&  \exp{\left[\lam_A(t-t_\act)\right]},
\label{LactLS} 
\EEA
where $t_\act$ is given by the condition $L_\act(t_\act) = 1$.
\paragraph*{High-specificity (HS) regime.}
In this regime, the specificity of activation is limited by the complexity of the naive repertoire. According to Eq.~(\ref{eq:R2}), the function $\Omega_\act$ is peaked around the moving front, $K_1(t)$ [see Fig.~\ref{fig:3}(b)]. In this case, we obtain 
\BEA
L_\act(t) &=&  \int \Omega_\act (K, t) \, \d(\log K)\qquad \mbox{(HS})\nonumber\\
& \approx  & \Omega_0 (K_1(t))\nonumber\\
&&\times \int\frac{\Omega_0(K) \, \d(\log K)}{(\Omega_0 (K_1(t))(1+K/K_\step)^p}
\label{Kint} \\
& = & \exp\left[\frac{\lam_A\beta^*}{p}(t-t_\act)\right].
\label{LactHS} 
\EEA
Here we use Eqs.~(\ref{eq:K_R}) and~(\ref{SE}), and we note that the integrand in (\ref{Kint}) has a peak value of order 1 and depends only weakly on $t$. 

These results are given in Eqs.~(\ref{eq:act1}) and~(\ref{eq:act2}) of the main text. In the activation dynamics discussed here, we assume that genetically distinct B cell lineages are also distinguishable in terms of their antigen binding affinity. Specifically, in our energy models, random mutations generate a binding energy change of order $k_B T$. If the sequence-energy map is highly degenerate, multiple activations occurring in sequence clusters of very similar antigen affinity can generate new scaling regimes. 

\vspace{1ex} \paragraph*{\bf Statistics of clone size.} 
Here we compute the cumulative distribution function (CDF) of clone size, $\Phi (N, t)$, which is defined as the fraction of activated clones with size $>N$ at time $t$. Given exponential growth with rate $\lambda_B$, this function is given by the fraction of lineages activated before a time $t' (N)$, 
\BE
\Phi (N, t) = \frac{L_\act (t'(N))}{L_\act (t)} 
\quad \mbox{with} \quad t' (N) = t - \frac{\log N}{\lambda_B}. 
\EE
Using Eqs.~(\ref{LactLS}) and~(\ref{LactHS}), we obtain 
\BE
\Phi(N, t) = \exp{[\zeta (t' (N) - t_\act)]} \sim N^{-1/\zeta}, 
\label{PhiN}
\EE
where the exponent $\zeta$ is defined in Eq.~(\ref{eq:zeta}) of the main text. In a similar way, we compute the expected size of the $j$-th largest clone, $\langle N_j \rangle (t)$ ($j = 1,2,3, \dots$). We write $\langle N_j \rangle (t) \sim \exp [\lam_B(t-t_j)]$, where the activation time $t_j$ is given by the condition $L_\act(t_j)=j$. Using again Eqs.~(\ref{LactLS}) and~(\ref{LactHS}), we have $t_j \sim \zeta \log j$ and 
\BE
\langle N_j \rangle (t) \sim j^{-\zeta} . 
\label{eq:N_j}
\EE
Eqs~(\ref{PhiN}) and~(\ref{eq:N_j}) are related by Zipf's law. Both are independent of  $t$ and, hence, valid also for the saturation clone sizes, $\bar N_j$, as used in Eq.~(\ref{eq:e_i}) of the main text. 

In the special case of a classical Luria-Delbr\"uck proliferation-mutation process, $L_\act (t)$ denotes the number of mutant clones present at time $t$, and Eqs.~(\ref{LactLS}) and (\ref{LactHS}) reduce to the simpler form $L_\act(t) = \exp (\lam_A t)$, where $\lambda_A$ is the wild-type growth rate. The resulting clone size statistics is still of the form (\ref{PhiN}) and (\ref{eq:N_j}) with size exponent $\zeta_0 = \lambda_B/\lambda_A$, where $\lambda_B$ is the mutant growth rate.

\vspace{1ex} \paragraph*{\bf Statistics of antigen affinity.} 
We now evaluate the CDF of antigen binding constants, $\Phi (K, t) = L_\act (K, t) / L_\act (t)$, in the HS regime. For the high-affinity tail of this function, $K \gtrsim K^*$, activation occurs deterministically, which implies $L_\act (K, t) \simeq L_0 (K)$. Recalling that $\Omega_0 (K) = (K \d / \d K) L_0 (K)$, we obtain $\Phi(K) \sim L_0 (K) \sim \Omega_0 (K)$. Hence,  
\BE
\Phi(K, t) \sim K^{\beta^*} \qquad \mbox{(HS}). 
\EE
Using Zipf's law, as for the clone size, we obtain the expectation value of the $l$-th lowest binding constant, 
\BE
\langle K_l \rangle (t)  \sim l^{1/\beta^*} \qquad \mbox{(HS}) , 
\label{eq:K_l}
\EE
as given in Eq.~(\ref{eq:e_i}) of the main text. Again, this relation is independent of $t$ and valid also at the saturation point. 
Because activation occurs on the moving front $K_1(t)$, the clone rankings by size and affinity are equivalent up to fluctuations. Hence, by combining Eqs. ~(\ref{eq:N_j}) and~(\ref{eq:K_l}), we obtain a power law relating size and affinity 
\BE
\langle \bar N_j\rangle \sim \langle K_j\rangle^{-\zeta\beta^*}  \qquad \mbox{(HS}).
\label{eq:N_K2}
\EE
In the LS regime, there is no clear power law relation between affinity and rank [see Fig.~\ref{fig:4}(b)]. High-affinity activated clones span the range between $K^*$ and $K_p$ and show a faster decline of affinity with rank than in the HS regime [see Figs.~\ref{fig:4}(b) and~\ref{fig:4}(c)]. Hence, empirical exponents fitted to affinity-rank data take values   $>1/\beta^*$ [see Fig.~\ref{fig:4}(d)].

\vspace{1ex} \paragraph*{\bf Potency statistics and elite neutralisers.} 
In the HS regime, Eqs. ~(\ref{eq:N_j}) and~(\ref{eq:K_l}) also determine the statistics of single-clone potencies $\bar z_j = \bar N_j /K_j$, 
\BE
\langle \bar z_j \rangle \sim j^{-\zeta -1/\beta^*} \qquad  \mbox{(HS}).
\EE
For typical individuals, many clones contribute to the total potency $\bar \Z=\sum \bar z_i$. However, a characteristic of Luria-Delbrück-Delbr\"uck models is the existence of giant fluctuations. In the HS regime of the model proposed here, there is a set of elite neutralisers singled out by early activation of their first clone. These ``jackpot'' clones have simultaneously high affinity and large size, which takes a sizeable fraction of the total activated repertoire, $\bar N_1 \lesssim \bar N$ [see Fig.~\ref{fig:6}(c)]. The CDF of potency, $\Phi (\Z)$, is defined as the fraction of individuals with $\bar \Z < \Z$. For $\bar \Z \gg \langle \bar \Z \rangle$, potency is dominated by jackpot clones, $\bar Z \simeq \bar z_1$. We find two scaling regimes [see Fig.~S3(c)]. In the pre-asymptotic regime ($\bar N_1 < \bar N$), size fluctuations of the jackpot clone are dominant, and we can write $\bar z_1 \approx \bar N_1/K^*$. Hence, by Eq. (\ref{PhiN}), the CDF of potency takes the form 
\BE
\Phi(\Z)\sim \Z^{-1/\zeta}  \qquad \mbox{(HS, pre-asymptotic regime)}.
\EE
In the asymptotic regime ($\bar N_1 \approx \bar N$), affinity fluctuations are dominant, and we have $\bar z_1 \approx \bar N/K$. Hence, 
\BE
\Phi(  \Z)\sim  L_0(\bar N/ \Z) \qquad  \mbox{(HS, asymptotic regime)}. 
\EE

\section{Empirical clone size statistics in early germinal centers} 
\label{app:C}
\vspace{1ex} \paragraph*{\bf Datasets of vaccination-induced B cell response.} 
We analyze sequencing data of early GCs from Refs.~\cite{Tas2016, Mesin2020} as a proxy for the initial population of activated B cell clones. We use CDR3 sequence counts to estimate the clone size of the different B cell lineages taking part in the response. The dataset contains samples from 10 independent lymph nodes 6 days post immunization with the model antigen chicken gamma globulin (CGG) (each lymph node belongs to a different mouse); 4 lymph nodes correspond to first immunization, 6 to secondary immunization. Given that participation of memory lineages is highly restricted, recall GCs are mainly seeded by de-novo recruited, naive B cell lineages~\cite{Mesin2020}. Data from immunization with other antigens are undersampled for our statistical analysis.

\vspace{1ex} \paragraph*{\bf Empirical statistics of clone size.} We calculate the expected scaled clone size of the $j$-th largest clone found in each lymph node, $N_j/ N_1$ [thin lines in Fig.~\ref{fig:4}(a), insert] and the average value over all 10 lymph nodes [thick line in Fig.~\ref{fig:4}(a), insert]. We obtain an empirical value of $\zeta$ by fitting these data to the form $N_j \sim j^{-\zeta}$, as given by Eq.~\ref{eq:x_i}. We also calculate the clonal entropy $\Sigma = -\sum_j x_j\log x_j$ with $x_j = N_j/N$ and $N=\sum_j N_j$ (the sum runs over all clones in the dataset). The empirical values of $\Sigma$ and $\zeta$ are marked in Fig.~S6 by circles with error bars. 

\vspace{1ex} \paragraph*{\bf Calibrated recognition model.} 
To compare the data with the model, we evaluate the effective antigen growth rate $\lambda^* (D, r_0, N_0)$, as given by Eq.~(\ref{lambdastar_vac2}). Using  the dosage used in the experiments, $N_0 \approx 10^{13}$, and physiological values of the diffusion constant, $D = 3\cdot 10^{-3}$~\cite{Beauchemin2006}, and the recognition radius, $r_0 = 0.5{\rm cm}$, we obtain $\lambda^* \approx 5.4 \, {\rm d}^{-1}$. Then we evaluate the repertoire entropy $\Sigma$, using stationary clone sizes $\bar N_j$, given by $\bar N_j \sim K_j^{-p}$ in the LS regime and by Eq.~(\ref{eq:N_K2}) in the HS regime, as well as the clone size exponent $\zeta$, given by Eq.~(\ref{eq:zeta}). We use physiological parameters $\lambda_B = 2 \, {\rm d}^{-1}$~\cite{Bocharov1994} and $\beta^* = 2.2$, computed as described in the theory section for a mouse repertoire size $L_0 = 10^8$~\cite{Altan-Bonnet2020}. Together, we obtain functions $\Sigma (p)$ and $\zeta (p)$ with $p$ as the only free parameter (Fig.~S6). For both observables, data and model are seen to be compatible for $p \approx 3$.

\section{Model parameters and numerical simulations} 
\label{app:D}

In the analytical and numerical analysis, we use the following empirical parameters for the activation process and the B cell repertoire:
(i) Size of the naive repertoire: $L_0 = 10^9$ lineages~\cite{Glanville2009, Morbach2010, Elhanati2015, Altan-Bonnet2020}. 
(ii) Growth rate of the antigen population: $\lam_A = 6 \text{d}^{-1}$~\cite{Smith2010, Pawelek2012, Goyal2021, Sender2021}.
(iii) Proliferation rate of activated B cells: $\lam_B = 2 \text{d}^{-1}$~\cite{Bocharov1994}.
(iv) Activation step rate: $k_\step = 0.5\text{min}^{-1}$. As discussed in the main text, this value is tuned to a repertoire size $L_0=10^9$ for humans and $L_0 = 10^8$ for mice~\cite{Altan-Bonnet2020}. 
(v) Antigen-BCR diffusion-limited association rate, $k_\on = 10^6 \text{M}^{-1}s^{-1}$~\cite{Pecht1972}.
(vi)~Carrying capacity of activated B cells: $\bar N=10^4$ cells \cite{Bocharov1994}.
In the optimality analysis of tuned repertoires [see Fig.~S5], we vary $p$, $k_{\rm step}$, and $L_0$ around the LS-HS crossover point given by $p \sim \beta^* (L_0)$, $k_\step \approx k_\on K^* (L_0)$. 

To simulate a primary B cell response, we start by creating an initial population of $L_0$ B cell lineages. Each B cell lineage is represented by a BCR sequence of length $\ell$, randomly drawn with the set of 20 amino acids. Assuming a deterministic expanding antigen concentration as defined in the main text, we calculate the non-homogeneous activation rate of each B cell lineage, $u_\act(t, K)$, as given in Eq.~\ref{eq:u_act}. Here we use the fact that the time for the first event, $t_1$, in a non-homogeneous Poisson process with rate $u(t)$ is distributed according to
\BE
P(t_1) = \frac{1}{{\cal N}} u(t_1)\exp\left[-\int_0^{t_1}u(t)dt\right].
\label{eq:times_sim}
\EE
with ${\cal N} = \int_0^\infty P(t)dt$. We sample then the activation time of each B cell lineage by sampling uniformly distributed random numbers $r\sim{\cal U} [0, 1]$ and then using the inverse of the cumulative version of Eq.~\ref{eq:times_sim}. 
Once we have all activation times, we proceed to determine the clone size of each of the B cell clones. Here we integrate $L_\act$ coupled differential equations 
\BE
\dot N_\b(t) = \lam_B N_\b(t) \left(1-\frac{\sum_\b N_\b(t)}{\bar N}\right),
\EE
assuming all clones start with clone size $N_\b(0) = 1$. 
We neglect all B cell clones whose final clone size is smaller than $\bar N_\b=2$, corresponding to less than one cell division.

\bibliography{references}

\newpage

\renewcommand\figurename{Fig.~S\!}
\setcounter{figure}{0}

\begin{figure*}[h!]
\centering
\includegraphics[width = .85\textwidth]{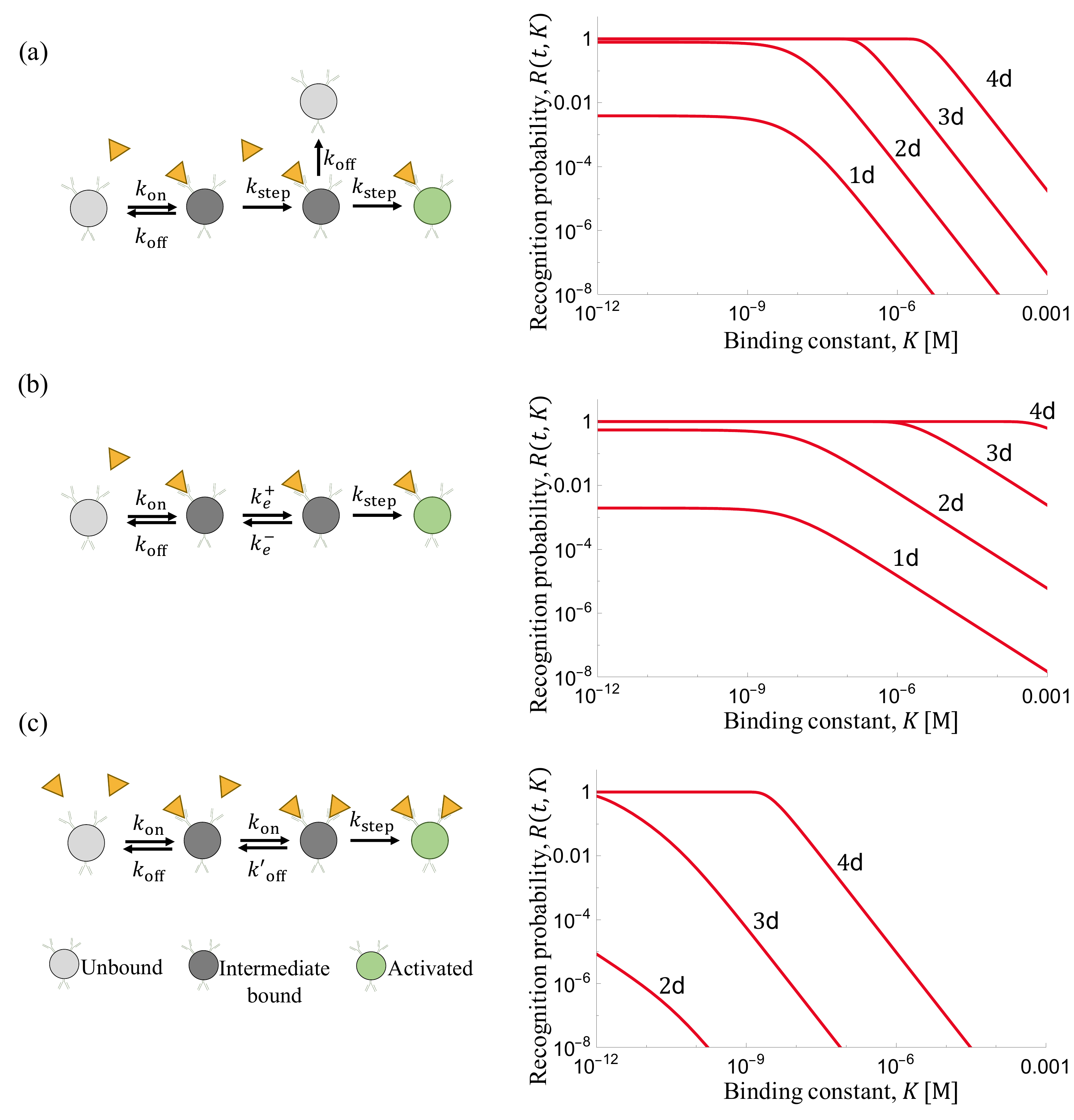}
\caption{\footnotesize {\bf Mechanisms of B-cell activation.}  
A schematic of the activation process (left) and the recognition probability $R(t, K)$ ($t = 1{\rm d}, \dots, 4{\rm d}$ after the start of the infection) are shown for the kinetic proofreading model discussed in the main text and for two models with reversible binding kinetics.
{\bf (a)} Activation by kinetic proofreading (as in Fig.~1 in the main text). This process has $p$ intermediate bound states and $p$ irreversible activation steps ($p >1$, shown here for $p = 2$). For antigen affinities of typical naive B-cells ($K \gtrsim 10^{-7}M$), kinetic proofreading generates fast, deterministic recognition of high-affinity lineages ($R \simeq 1$) and strong suppression of low-affinity lineages ($R \sim K^{-p}$ for $K \gg K_1(t)$). 
{\bf (b)} Activation via an intermediate excited state. This process is similar to the kinetic proofreading model, but the binding kinetics is reversible. It generates fast, deterministic of high-affinity lineages but only weak suppression of low-affinity lineages ($R \sim K^{-1}$ for $K \gg K_1(t)$). 
{\bf (c)} Activation by cooperative binding. In this process, activation requires simultaneous binding of $p$ virions to the same B-cell ($p >1$, shown here for $p = 2$). It generates strong suppression of low-affinity lineages ($R \sim K^{-p}$ for $K \gg K_\step$), but even high-affinity naive lineages do not reach deterministic activation ($R \ll 1$ for $K \sim 10^{-7}M$). 
Kinetic parameters: $k_\on=10^6 s^{-1}M^{-1}$; $\delta = 10^{-2}$ ; $k_\step = 0.5\text{min}^{-1}$; $k_\step^-=0.1k_\step$.
} 
\label{fig:S1}
\end{figure*}
\newpage

\begin{figure*}[h!]
\centering
\includegraphics[width=.9\textwidth]{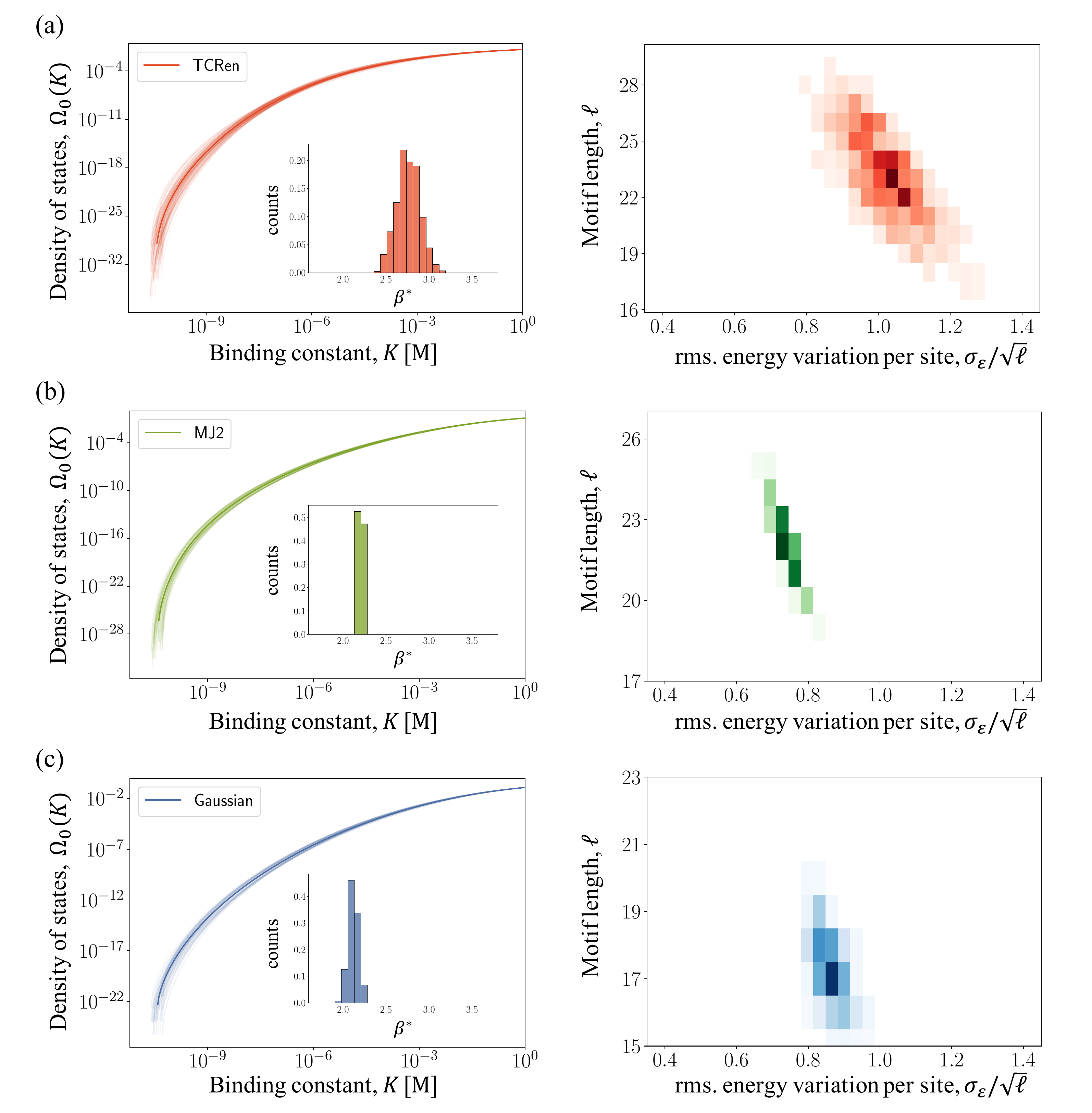}
\caption{\footnotesize {\bf Density of states in different energy models. }  
Left: density of naive B-cell lineages, $\Omega_0 (K)$, for a sample of antigen epitope sequences with $K_m \approx 10^{-11} M$  and $K^* \approx 10^{-7} M$ in a repertoire of overall size $L_0 = 10^9$ (thin lines: individual antigens, thick line: ensemble average); see Appendix~B. Insert: ensemble distribution of the inverse activation temperature $\beta^*$, as defined in the main text. Right: joint distribution of the binding length, $\ell$, and of the rms. energy variation per site, $\sigma_\epsilon / \ell^{1/2}$. Binding motifs are randomly sampled from different additive energy models: 
{\bf (a)} TCRen matrix~\cite{Karnaukhov2022}, 
{\bf (b)} Miyazawa-Jernigan matrix~\cite{Miyazawa1996}, 
{\bf (c)} Gaussian random energies; details are given in Appendix~B.
Model parameters as in Fig.~2.}
\label{fig:S2}
\end{figure*}
\newpage

\begin{figure*}[ht!]
\centering
\includegraphics[width= .95\textwidth]{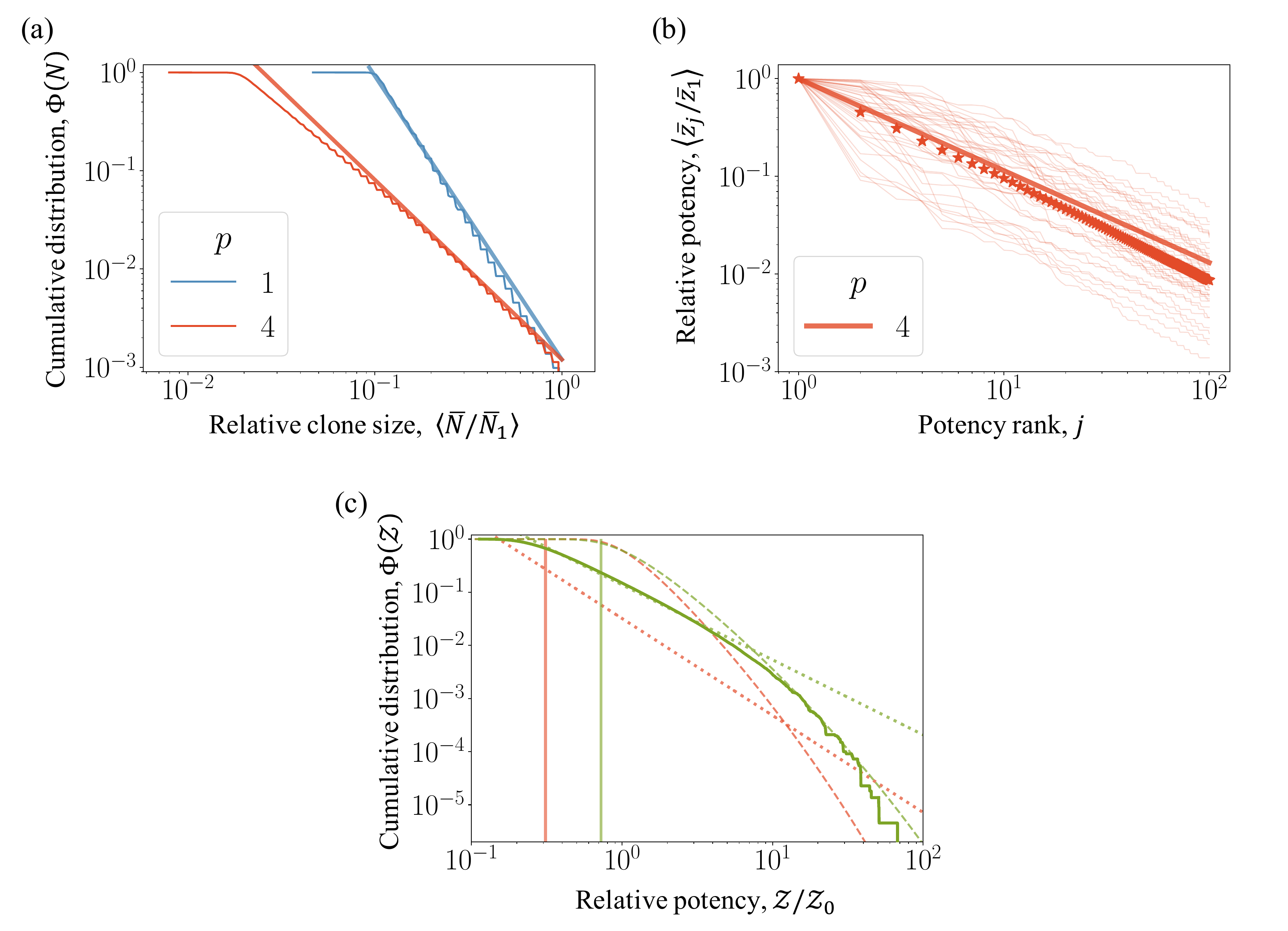} \\
\caption{\footnotesize {\bf Statistics of activated repertoires.}
{\bf (a)}  Cumulative clone size distribution aggregated over individuals, $\Phi (N) \sim N^{-1/\zeta}$ (blue: LS regime, $p = 1$; red: HS regime, $p = 3$). 
{\bf (b)} Potency-rank relation in the HS regime,  $\langle \bar z_j \rangle \sim j^{-\zeta -1/\beta^*}$ for potency-ranked clones ($j = 1, \dots, 100$). Thin lines give rankings in a set of randomly chosen individuals.  
{\bf (c)} Cumulative population distribution function of potency in the HS regime, $\Phi (\Z)$ ($p=4$; $L_0=10^7$ in green and $L_0=10^9$ in red). Elite neutralisers with jackpot clones define a pre-asymptotic regime, $\Phi (\Z) \sim \Z^{-1/ \zeta}$ (dotted), and an asymptotic regime, $\Phi (\Z) \sim L_0 (\bar N / \Z)$ (dashed). Vertical lines indicate population average $\langle \Zbar \rangle / \langle \Z_0 \rangle$. Model parameters as in Fig.~2 in the main text. 
}
\label{fig:S3}
\end{figure*}
\newpage

\begin{figure*}
\centering
\includegraphics[width= 0.95 \textwidth]{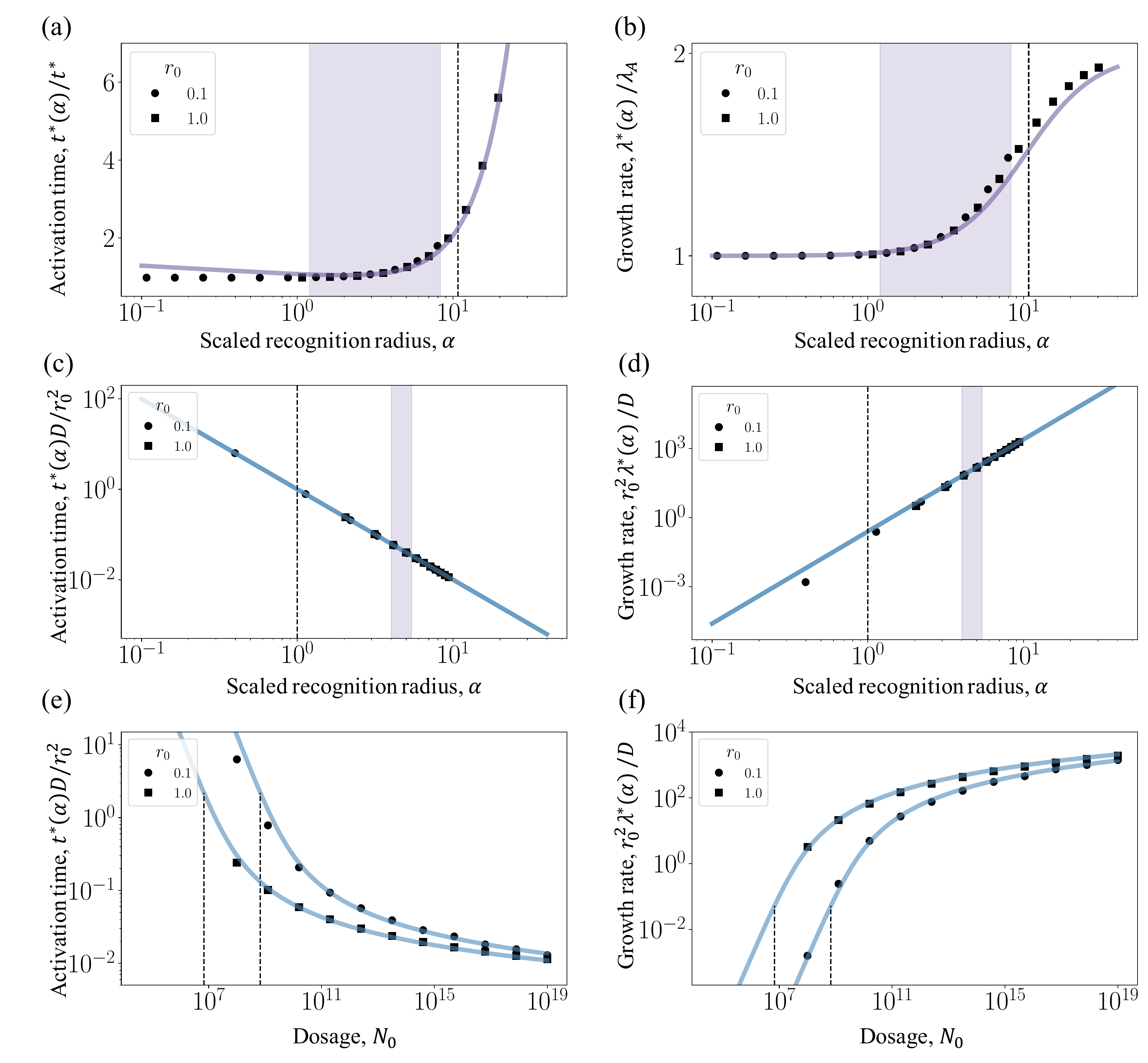}

\caption{\footnotesize {\bf Onset of B cell activation in spatio-temporal processes.} 
{\bf (a,b)} Scaled onset time $t^* (D, r_0)/t^* = g_{t, {\rm inf}} (\alpha)$ and scaled effective growth rate, $ \lambda^* (D, r_0)/\lambda_A = g_{\lambda, {\rm inf}} (\alpha)$ , for the case of an infection, where $\alpha$ is varied by variation of $D$ for given values of $r_0$.
{\bf (c,d)}~Scaled onset time $t^* (D, r_0)/(r_0^2/D) = \alpha^{-2}$ and scaled effective growth rate, $\lambda^* (D, r_0)/(D/r_0^2) = g_{\lambda, {\rm vac}} (\alpha)$, for the case of a vaccination, where $\alpha$ is varied by variation of $N_0$ for given values of $r_0$ and $D = 3\cdot10^{-3}{\rm cm}^{2}{\rm d}^{-1}$. Theoretical values of these scaling functions, given by Eqs.~(\ref{tstar_inf}), (\ref{lambdastar_inf}), (\ref{eq:t*vac}), and (\ref{lambdastar_vac2}), are shown together with numerical evaluations of the onset condition (\ref{onset_st}). 
{\bf (e,f)} Dosage dependence of onset time and effective growth rate. 
Shading marks the regime of physiological parameters, a dashed line the cross-over scale $\tilde \alpha$ from weak to strong diffusion constraint.
Other model parameters as in Fig.~\ref{fig:2}.
}
\label{fig:S4}
\end{figure*}

\vspace*{3cm}

\begin{figure*}[ht!]
\centering
\includegraphics[width= .9\textwidth]{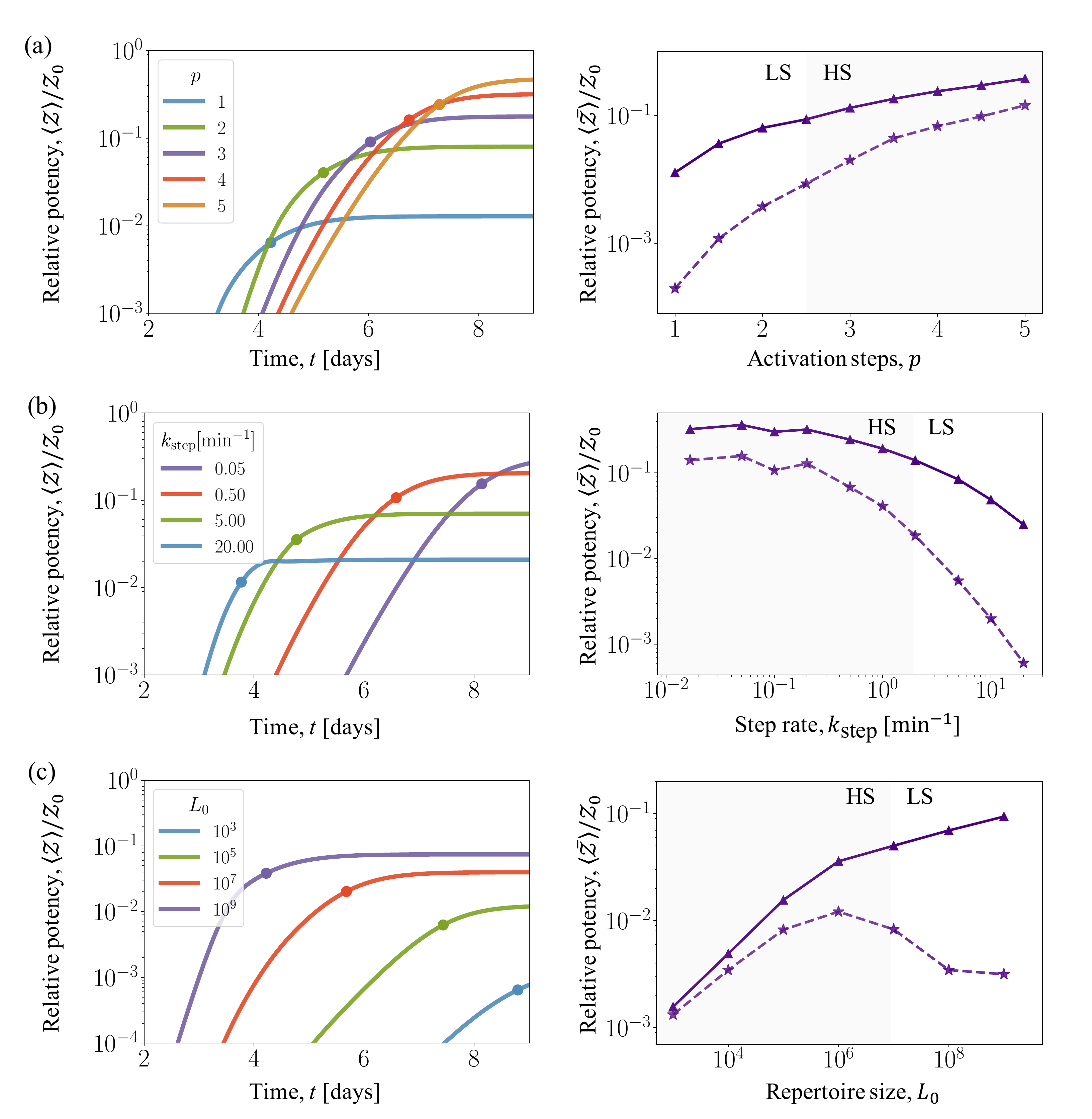}
\caption{\footnotesize {\bf Optimality of antigen recognition by tuned proofreading.} 
Antiserum potency is shown for variation of the proofreading parameters $p$, $k_\step$, and of the repertoire size, $L_0$, around the LS-HS crossover point ($p \sim \beta^* (L_0)$, $k_\step \approx k_\on K^* (L_0)$, $L_0 = 10^9$), which characterises a tuned repertoire as discussed in the main text. 
Left: time-dependent, population averaged potency, $\Z (t)$; the half-saturation point $(t_{50}, Z_{50})$ is marked by a dot. 
Right: saturation value, $\Zbar$ (thick lines), and the potency component of the largest clone, $\bar z_1$ (dashed lines). All potencies are shown in units of the reference value $\Z_0$ defined in the main text; the HS regime is marked by shading.
{\bf (a)} Variation of the number of proofreading steps, $p$, at constant $k_\step$ and $L_0$ (cf.~Figs.~6(a) and 6(b) in the main text). 
In the LS regime ($p \lesssim \beta^*$), potency increases strongly with $p$. In the HS regime ($p \gtrsim \beta^*$), increasing $p$ yields a diminishing return of potency, while $t_{50}$ continues to increase proportionally to $p$.
{\bf (b)} Variation of the rate of proofreading steps, $k_\step$, at constant $p$ and $L_0$. 
In the LS regime ($k_\step \gtrsim k_\on K^*$), potency increases strongly with decreasing $k_\step$. In the HS regime ($k_\step \lesssim k_\on K^*$), decreasing $k_\step$ yields a diminishing return of potency, while $t_{50}$ continues to increase proportionally to $1/ k_\step$.
{\bf (c)} Variation of the repertoire size, $L_0$, at constant $p$ and $k_\step$. Here we use $p=2$ and $k_\step = 
0.05 \text{min}^{-1}$, which are tuned for $L_0=10^7$.
In the HS regime ($L_0 \lesssim 10^{7}$), potency increases strongly with $L_0$. In the LS regime ($L_0 \gtrsim 10^{7}$), increasing $L_0$ yields a diminishing return of potency and the potency component of the largest clone decreases.
Other model parameters as in Fig.~2 in the main text.
}
\label{fig:S5}
\end{figure*}
\newpage

\begin{figure*}
\centering
\includegraphics[width= 1\textwidth]{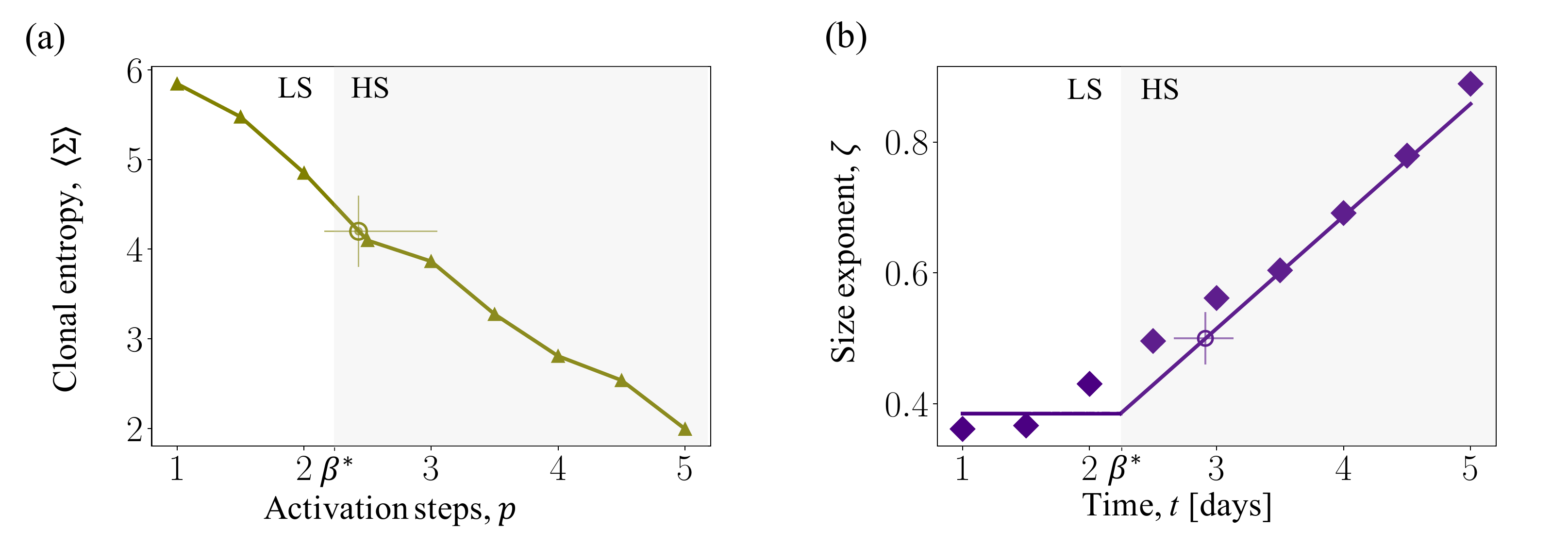}
\caption{ \footnotesize {\bf Analysis of mouse repertoire data.} 
Model predictions for the repertoire entropy, $\langle \Sigma \rangle$, and the clone size exponent, $\zeta$, as functions of the proofreading order, $p$, are compared with empirical values inferred from the clone size data of refs.~\cite{Tas2016, Mesin2020} (open circles with error bars). 
Parameters: repertoire size $L_0 = 10^{-8}$, diffusion constant $D=3\cdot10^{-3}\mathrm{cm}^2\mathrm{d}^{-1}$, recognition radius $r_0 = 5 \mathrm{mm}$; other parameters as in Fig.~\ref{fig:2}. The inference procedure is detailed in Appendix~C.
\label{fig:S6}
}
\end{figure*}
\newpage
\end{document}